\documentclass[amsmath,floats,floatfix,12pt,nofootinbib,tightenlines,showpacs]{revtex4}

\usepackage{graphicx}
\usepackage{amssymb}
\usepackage{bm}

\newcommand{\shift}{V}
\newcommand{\Binner}{{\mathcal{B}_i}}
\newcommand{\Bouter}{{\mathcal{B}_o}}

\begin{document}
\title{IMEX evolution of scalar fields on curved backgrounds}

      \author{Stephen~R.~Lau${}^{1,}$\footnote{
              {\tt lau@dam.brown.edu}, 
              ${}^\ddagger\;{}${\tt harald@tapir.caltech.edu},
              ${}^\S\,{}${\tt Jan\_Hesthaven@brown.edu}}${}^{,}$\footnote{
              Also at the
              Department of Mathematics \& Statistics,
              University of New Mexico,
              Albuquerque, NM 87131.},
              Harald~P.~Pfeiffer${}^{2,\ddagger}$, and
              Jan~S.~Hesthaven${}^{1,\S}$}
\address{${}^1$Division of Applied Mathematics, Brown
               University, Providence, RI 02912.\\
         ${}^2$Theoretical Astrophysics and Relativity
               Group 130-33, California Institute of
               Technology, Pasadena, CA 91125.}
\begin{abstract}
Inspiral of binary black holes occurs over a 
time-scale of many orbits, far longer than the dynamical 
time-scale of the individual black holes.  Explicit 
evolutions of a binary system therefore require excessively many 
time-steps  to capture interesting 
dynamics. We present a strategy to overcome the Courant-Friedrichs-Lewy 
condition in such evolutions, one relying on modern implicit-explicit ODE 
solvers and multidomain spectral methods for elliptic equations. Our analysis  
considers the model problem of a forced scalar field propagating on a 
generic curved background. Nevertheless, we encounter and address a number 
of issues pertinent to the binary black hole problem in full general
relativity. Specializing to the Schwarzschild geometry in Kerr-Schild
coordinates, we document the results of several numerical experiments 
testing our strategy.
\end{abstract}

\pacs{04.25.Dm, 02.70.Hm; AMS numbers: 65M70, 83-08, 83C57}

\maketitle
\section{Introduction}
\label{sec:intro}

Numerical simulations of the inspiral and merger of binary black holes 
(BBH) investigate Einstein's equations in the nonlinear regime where 
analytical progress often proves intractable. 
The primary goal of these simulations is the computation 
of gravitational waveforms necessary to analyze output from 
gravitational wave detectors like the 
``Laser Interferometric Gravitational Wave Observatory'' (LIGO). 
Breakthroughs in 2005 have yielded two ways to simulate BBH evolutions: 
the {\em generalized harmonic system} (GHS) with excision~\cite{Pretorius2005} 
and the Baumgarte-Shapiro-Shibata-Nakamura (BSSN) system with 
{\em moving punctures} \cite{UTB2006,NASA2006}. Over the last few 
years numerical relativity has seen rapid progress along both fronts.

The evolution of a binary black hole proceeds through three phases. 
During the {\em inspiral} phase, the two separate black holes orbit about 
each other, with the orbit gradually tightening due to emission of angular 
momentum and energy via gravitational radiation. At small separation, the black 
holes encounter a dynamical instability, plunge rapidly toward each other and 
merge. This {\em merger} phase results in a single, larger, highly distorted 
black hole which subsequently relaxes to a stationary black hole during 
the {\em ringdown} phase.  Merger and ringdown happen quickly, together 
lasting about $200M$, where the black hole mass $M$ sets both the spatial 
and temporal scales.  Therefore, merger and ringdown are comparatively easy 
to simulate at modest computational cost. In contrast, simulation of the 
inspiral phase is a daunting computational challenge. Because the orbital 
period increases rapidly with separation of the black holes, simulation of 
even a modest number of orbits requires much longer evolutions. For example, 
the last 10 orbits of an equal mass non-spinning binary black hole last about 
$2000M$, already an order of magnitude longer than merger and ringdown. 
Beyond necessarily longer time-spans, inspiral simulations also require 
higher accuracy. Indeed, gravitational wave flux decreases with separation, 
and it must be accurately resolved in order to compute the correct phasing 
of the gravitational waves. 

To date all binary black hole simulations have employed explicit 
time-stepping, generally the method of lines with an explicit ODE scheme 
like the classical fourth-order Runge-Kutta method.  Without question 
explicit time-stepping is appropriate for both merger and ringdown. 
However, during the inspiral phase, the relevant physical time-scale 
on which the binary separation changes is much longer than the dynamical 
time-scale $M$ of each black hole. Nevertheless, the Courant condition 
associated with an explicit time-stepper heuristically requires that 
time-steps are proportional to the smallest grid spacing, and 
therefore explicit binary evolutions use time-steps that are typically 
of the order $M/100$ to $M/10$. For instance, a recent 16 orbit 
simulation~\cite{Boyle2007} required nearly 200,000 
explicit time-steps. This issue becomes
more pronounced when modeling black holes with unequal masses $M_1>M_2$. 
The orbital period is proportional to the total mass $M=M_1+M_2$, whereas 
the Courant limit dictates that the time-step is proportional to the smaller 
mass $M_2$.  The number of explicit time-steps needed to ensure numerical 
stability then scales like $M/M_2$.  Due to these reasons,
only a few binary black hole simulations with mass ratios above 4:1 
have been performed~\cite{Baker2008,Sperhake2008}, 
and these are quite short and computationally expensive. 
Courant limitations are likewise more 
severe for simulations of spinning black holes, which also require 
higher spatial resolution close to the black holes.

These arguments suggest that some form of implicit time-stepping would
more efficiently treat the inspiral phase, and this paper begins the study 
of alternative ways to carry out temporal integration of orbiting binaries 
in the early phase of their evolution.  Our approach is based on modern 
implicit-explicit (IMEX) ODE solvers~\cite{KC-ARK,Dutt2000,Minion2003} and 
classical multidomain spectral methods\footnote{The IMEX strategy is more 
general, and, for example, is also applicable to discontinuous Galerkin 
methods \cite{Kanevskyetal2007}.}, and in particular 
those~\cite{Pfeiffer2003} used for solving elliptic problems (it turns out 
that our implicit equations correspond to elliptic PDE).  The generalized 
harmonic formulation~\cite{Pretorius2005c,Pretorius2006,Lindblom2006} 
rewrites Einstein's equations as 10 scalar wave equations for 
the components of the metric, which are coupled through nonlinear lower 
order terms.  Because the principal part is just the scalar wave operator 
on a curved background spacetime, we consider here the model problem of 
evolving a scalar field on a single black-hole background.
IMEX schemes like the ones pursued here are not the only possible 
approach to circumvent the Courant-Friedrichs-Lewy condition.  
For instance, Hennig and Ansorg~\cite{HennigAnsorg2008} explore 
spacetime spectral methods to solve scalar wave equations.

The organization of the paper is as follows. The upcoming Section 
\ref{sec:prelim} gives a brief 
overview of IMEX methods, using the specific example of Additive Runge-Kutta 
(ARK). It also briefly collects the relevant first-order equations 
describing the propagation of scalar waves on a generic curved spacetime,
and discusses boundary conditions for such equations. Section 
\ref{sec:impequations} contains our main analytical discussion, and it
focuses on the novelty of solving the implicit equations which arise in 
our time-stepping strategy. Much of this theoretical analysis is general, 
but we eventually settle on the concrete example of a scalar field propagating 
on the Schwarzschild geometry in Kerr-Schild coordinates. Section 
\ref{sec:numtests} describes the results of several numerical experiments 
carried out for the Schwarzschild scenario.  The conclusion in
  Sec.~\ref{sec:conclusion} summarizes our findings and discusses
  steps necessary for application of IMEX methods to the ultimate
  target problem, binary black hole inspiral.  Finally, three appendices
collect some technical calculations omitted in Section
\ref{sec:impequations}.

\section{Preliminaries}
\label{sec:prelim}
\subsection{Implicit-explicit additive Runge-Kutta}
\label{subsec:IMEX-ARK}
From the computational point of view, all IMEX methods require 
that we are able to numerically solve an implicit equation. For 
concreteness, we here consider ARK3(2) and ARK4(3), two 
IMEX additive Runge-Kutta schemes introduced in \cite{KC-ARK}. 
These schemes share the same algorithmic 
structure (only their sets of Butcher tableaux differ).
ARK3(2) is a 4-stage third-order scheme with a 
second-order embedded scheme, while ARK4(3) is a 6-stage 
fourth-order scheme with a third-order embedded scheme.
Although we will not report on it 
here, we have also considered various versions of 
{\em semi-implicit spectral deferred corrections} (SISDC)
\cite{Dutt2000,Minion2003} as an alternative to ARK. The 
nature of the SISDC algorithm is quite different, but its 
implementation also requires that we are able to solve 
(in this case at each substep) an implicit equation of 
the same form.

We will not discuss accuracy and stability properties of ARK. 
Our purpose here is simply to describe the algorithm, highlight 
what is needed for implementation, and focus on the origin and 
structure of the implicit equation. When considering first-order 
systems for scalar wave propagation below, we will adopt what is 
essentially a reversed semidiscrete picture. That is to say, we 
consider time as discrete, but retain the spatial continuum.
When adopting that picture, we will write down a continuum 
implicit equation (a spatial differential equation) that 
corresponds to the implicit equation appearing in the ARK algorithm. 
Although this should not prove cause for confusion, we have 
nevertheless raised this issue now, since we adopt a similar 
notation whether or not the spatial continuum is retained.

Mostly adopting the notation of \cite{KC-ARK},
we begin with a generic initial value problem
\begin{equation}\label{eq:GeneralEq}
\frac{d\boldsymbol{u}}{dt} =
\boldsymbol{f}(t,\boldsymbol{u})
= \sum_{\nu = 1}^2
\boldsymbol{f}^{[\nu]}(t,\boldsymbol{u})
,\qquad
\boldsymbol{u}(t_0) = \boldsymbol{u}_0,
\end{equation}
with $\boldsymbol{u}$ a vector of unknowns. Adopting a $2$-additive
scheme, we have split the right-hand side $\boldsymbol{f}$
into explicit (nonstiff) $\boldsymbol{f}^E = \boldsymbol{f}^{[1]}$ 
and
implicit (stiff) $\boldsymbol{f}^I = \boldsymbol{f}^{[2]}$ sectors.
The ARK schemes specify a rule for advancing the vector
$\boldsymbol{u}^n$ at a present time-step $t^n$
(perhaps $t_0$) to the vector $\boldsymbol{u}^{n+1}$ at the
next time-step $t^{n+1} = t^{n} + \Delta t$, and this rule
requires the construction of $s$ stage values $\boldsymbol{u}^{(i)}$,
$i=1, 2, \ldots, s$,
corresponding to intermediate times $t^{(i)} = t^{n} +
c_i \Delta t$. The first stage is given by $\boldsymbol{u}^{(1)} = 
\boldsymbol{u}^{n}$, and the
 remaining stage values are determined sequentially by
\begin{align}
\boldsymbol{u}^{(i)}
=
\boldsymbol{u}^{n}
+ \Delta t \sum_{j=1}^{i} \left[
a^E_{ij} \boldsymbol{f}^E(t^{(j)},\boldsymbol{u}^{(j)})
+ a^I_{ij} \boldsymbol{f}^I(t^{(j)}, \boldsymbol{u}^{(j)})\right]
,\quad 2 \le i \leq s.
\label{eq:uidef}
\end{align}
After all the stages have been computed,
the updated solution is given by the stage expansion
\begin{align}
\boldsymbol{u}^{n+1} & = \boldsymbol{u}^{n}
+\Delta t \sum_{i=1}^s b_i\left[
\boldsymbol{f}^E(t^{(i)}, \boldsymbol{u}^{(i)}) +
\boldsymbol{f}^I(t^{(i)}, \boldsymbol{u}^{(i)})\right].
\label{eq:unp1}
\end{align}
The parameter matrices $\mathbf{A}^\mathrm{ERK} = (a^E_{ij})$ and
$\mathbf{A}^\mathrm{ESDIRK} = (a^I_{ij})$, along with the 
coefficients
$\mathbf{b} = (b_i)$ and $\mathbf{c} = (c_i)$, stem from Butcher 
tableaux
collected in \cite{KC-ARK}.  ERK stands for {\em explicit 
Runge-Kutta},
and ESDIRK for {\em explicit singly diagonally implicit 
Runge-Kutta}.
In the ESDIRK acronym, the {\em explicit} refers to the
trivial first stage, and {\em diagonally implicit} to the fact that
the sum in (\ref{eq:uidef}) stops at $i$ rather than $s$.

For the explicit sector, $a^E_{ij}=0$ for $j\ge i$.
Therefore, in Eq.~(\ref{eq:uidef}) the first term in the sum 
depends solely on already known stages, $\boldsymbol{u}^{(1)}, \ldots,
\boldsymbol{u}^{(i-1)}$.  In contrast, for the implicit
sector, $a^I_{ij}=0$ for $j > i$, with $a^I_{ii} = \gamma \neq 0$
unless $i = 1$ (the {\em singly} in ESDIRK indicates that the
diagonal elements $a^I_{22}=a^I_{33}=\ldots=a^I_{ss}$ all equal 
the same constant $\gamma$).
The term $a_{ii}^I\boldsymbol{f}^I(t^{(i)},\boldsymbol{u}^{(i)})$
turns Eq.~(\ref{eq:uidef}) into an implicit equation for
$\boldsymbol{u}^{(i)}$.
Implementation of an ARK scheme therefore requires that we
are able to solve (at each stage after the first) an
{\em implicit equation} of form
\begin{equation}
\boldsymbol{u} -
\alpha \boldsymbol{f}^I(t, \boldsymbol{u})
= \boldsymbol{B},
\label{eq:abstractimpeqn}
\end{equation}
where $\alpha=\gamma\Delta t$ and 
$\boldsymbol{B}$ depends on the previous stage values.

\subsection{First-order equations for a scalar field on a 
curved background} \label{subsec:FirstOrderScalarWave}

Our goal is to solve the scalar wave equation
\begin{equation}\label{eq:ScalarWave2ndOrder}
\nabla_\mu\nabla^\mu\psi = S 
\end{equation}
with a given source term $S=S(t,x^k)$. We consider 
Eq.~(\ref{eq:ScalarWave2ndOrder}) on a generic curved background 
with line-element given in the usual 3+1 decomposition,
\begin{equation}
ds^2 = -N^2 dt^2 + g_{jk}
\left(dx^j+\shift^j dt\right)
\left(dx^k+\shift^k dt\right).
\end{equation}
Here $g_{jk}$ is the induced metric on $t=\mbox{const}$
hypersurfaces, $N$ is the lapse function, and 
$\shift^k$ is the shift vector.  
These quantities are known functions of space $x^j$ and
time $t$, where lower-case Latin indices $(j,k,\ldots)$ 
denote spatial components running over $1,2,3$.

Following Holst {\em et al.}~\cite{Holst2004}, we 
rewrite Eq.~(\ref{eq:ScalarWave2ndOrder}) as the
following first-order system:
\begin{subequations}\label{eq:FirstOrderScalarWave}
\begin{align}
\partial_t\psi & = \shift^k \partial_k \psi
- N\Pi
\label{eq:psidot}
\\
\partial_t \Pi & = \shift^k \partial_k \Pi
- N g^{jk} \partial_j \Phi_k
+ N K \Pi
+ N \Phi_k J^k
+ N S
\label{eq:Pidot}
\\
\partial_t \Phi_j & =
  \shift^k \partial_k \Phi_j
- N \partial_j \Pi
+ \Phi_k \partial_j  \shift^k
- \Pi \partial_j N.
\label{eq:Phidot}
\end{align}
\end{subequations}
Apart from the possible inhomogeneous forcing term $NS$ in
(\ref{eq:Pidot}), these are Eqs.~(14--16) of \cite{Holst2004}. 
$\Pi$ is a new evolved variable representing the
time derivative of $\psi$, and Eq.~(\ref{eq:psidot}) is 
the definition of $\Pi$. The $\Phi_j\equiv\partial_j\psi$
represent the spatial derivatives of $\psi$.  
The quantities $J^k$ and $K$ depend only on the background 
spacetime,
\begin{align}
J^k & = -N^{-1}g^{-1/2}\partial_j(Ng^{1/2} g^{jk})
\label{eq:Jk}\\
K & = -N^{-1}g^{-1/2}\big[\partial_t g^{1/2} - \partial_j
(g^{1/2}\shift^j)\big],
\label{eq:K}
\end{align}
where $K$ is the trace of the extrinsic curvature tensor and $g=\det(g_{jk})$.
These formulas for $J^k$ and $K$ are respectively 
Eqs.~(17) and (18) of \cite{Holst2004}.  

Solutions of the first-order system
Eqs.~(\ref{eq:FirstOrderScalarWave}) are equivalent to those of
Eq.~(\ref{eq:ScalarWave2ndOrder}) only if the constraint
\begin{equation}\label{eq:C}
{\mathcal C}_k\equiv \partial_k\psi-\Phi_k
\end{equation}
vanishes.  The constraint provides an important link between
  Eq.~(\ref{eq:psidot}) and (\ref{eq:Phidot}), as the latter equation
is derived by taking the time derivative of the constraint: 
\begin{equation}\label{eq:dt_C}
\partial_t\mathcal{C}_k=0\quad\Rightarrow\quad
\partial_t\Phi_k = \partial_t\partial_k\psi=\partial_k \partial_t\psi.
\end{equation}
Thus, the right-hand side of Eq.~(\ref{eq:Phidot}) is the gradient of
the right-hand side of Eq.~(\ref{eq:psidot}), with 
$\partial_k\psi$ replaced by $\Phi_k$.  

Boundary conditions relative to a boundary element with
outward-pointing unit normal $n_k$ are described in terms of the
characteristic fields
\begin{equation}
Z^{1} = \psi,\qquad Z^{2}_j = P^k_j \Phi_k,\qquad 
U^{1\pm} = \Pi \pm n^k \Phi_k,
\label{eq:charvars}
\end{equation}
where $P^k_j = g^k_j - n^k n_j$ indicates projection 
tangential to the boundary element. 
Relative to the time axis $\partial/\partial t$, 
the coordinate speeds of these fields are respectively
\begin{equation}
-n_k \shift^k,\qquad -n_k \shift^k,\qquad -n_k \shift^k \pm N.
\label{eq:charspeeds}
\end{equation}
A characteristic field requires a boundary condition whenever
its characteristic speed is negative. The scenario we consider 
later, that is wave propagation on a Schwarzschild black hole, 
has two boundaries: First, an outer spherical boundary 
$\Bouter$ where $Z^1$, $Z^2_j$ and $U^{1-}$ 
are incoming and require boundary conditions. Second, an inner 
spherical boundary $\Binner$ which is
inside the black hole horizon and surrounds
the singularity at the center of the black hole.  
On the inner boundary all characteristic 
fields are outgoing (i.~e.~moving toward the center of the black 
hole), and boundary conditions must not be
 imposed on it. This pure outflow boundary 
results in several interesting features of the present work to 
be discussed below. If $\Bouter$ and $\Binner$ are
adapted to the background symmetry (i.~e.~round spheres,
which they need not be for our numerical work), then  
$n^k\partial/\partial x^k \propto \partial/\partial r$ on 
$\mathcal{B}_o$ and $n^k\partial/\partial
x^k \propto -\partial/\partial r$ on $\mathcal{B}_i$.

The boundary condition on $U^{1-}$ is physical; this boundary 
condition and the choice of initial data  determines which 
solution of the second-order wave 
equation~(\ref{eq:ScalarWave2ndOrder}) is computed. In this 
paper, we typically choose the initial data, the boundary 
values of $U^{1-}$, and the external forcing $S$ such that 
the solution follows a prescribed exact solution. Boundary
conditions on the fields $Z^1$ and $Z^2_j$, if necessary, are chosen
to ensure that the constraint ${\cal C}_k$ vanishes on the boundary.
Solutions to the first-order system (\ref{eq:FirstOrderScalarWave}) 
which violate the constraint ${\cal C}_k = 0$ are not admitted either 
by the scalar equation (\ref{eq:ScalarWave2ndOrder}) or the reduced
system which arises from setting $\Phi_k = \partial_k\psi$ in 
(\ref{eq:FirstOrderScalarWave}). The boundary conditions on $Z^1$ and 
$Z^2_j$ rule out these spurious solutions to the extended system 
(\ref{eq:FirstOrderScalarWave}), provided that the initial data also
satisfies the constraint.  Such 
{\em constraint preserving boundary conditions}
have been derived by Holst et al.~\cite{Holst2004}
and refined by  
Lindblom et al.~\cite{Lindblom2006}. The implicit equations
that we encounter in our use of
ARK methods require boundary conditions that parallel those of 
the evolution problem.  We will therefore consider $U^{1-}$ as given 
boundary data for the implicit problems, as well as the boundary data
${\cal C}_k=0$, whenever necessary.

\section{Implicit equations}\label{sec:impequations}

\subsection{First-order equations}
The  ARK algorithm described in Sec.~\ref{subsec:IMEX-ARK}
is applicable only to systems of ODE, and for the case 
at hand such an ODE system arises upon spatial approximation of 
Eqs.~(\ref{eq:FirstOrderScalarWave}) via a pseudospectral collocation 
method (see, for example, \cite{Holst2004} for details). However, 
as mentioned earlier, we find it convenient to retain the spatial 
continuum in our discussion, and so write down the continuum 
implicit equations (PDEs) which, upon spatial approximation, yield 
the relevant algebraic implicit equations appearing in our IMEX 
algorithms. Equations~(\ref{eq:FirstOrderScalarWave}) have the form 
of Eq.~(\ref{eq:GeneralEq}) for the evolved variables 
$\boldsymbol{u}=(\psi,\Pi,\Phi_k)$.  
We will consider a number of possibilities for splitting the 
right-hand side of Eqs.~(\ref{eq:FirstOrderScalarWave}) into 
stiff (implicit) $\boldsymbol{f}^I$ and nonstiff 
(explicit) $\boldsymbol{f}^E$ sectors, 
but always treat the system's principal part 
(i.~e.~all spatial derivatives) implicitly.

Each of our possible choices for the 
IMEX splitting is specified by writing down 
the field components of the implicit 
equation (\ref{eq:abstractimpeqn}). Treating implicitly the first 
two terms from each right-hand side in~(\ref{eq:FirstOrderScalarWave}), 
and possibly the forcing term $NS$ 
from (\ref{eq:Pidot}), we get {\bf case (i)}:
\begin{subequations}\label{eq:case1}
\begin{align}
\psi - \alpha \left(
\shift^m \partial_m \psi
-N\Pi\right) & = B_\psi
\label{eq:psiimp1}\\
\Pi - \alpha \left(
\shift^m \partial_m \Pi
- N g^{jm} \partial_j \Phi_m + \epsilon N S\right) & = B_\Pi
\label{eq:Piimp1}\\
\Phi_k - \alpha \left(
\shift^m \partial_m \Phi_k
- N \partial_k \Pi
\right) & = B_{\Phi_k},
\label{eq:Phiimp1}
\end{align}
\end{subequations}
with $\epsilon = 1$ for implicit treatment of $NS$, and  $\epsilon = 0$ 
otherwise. Therefore, as with the 
other cases to follow, case (i) is actually two cases.
A second, and similar, set of equations stems from also 
treating implicitly all terms in the right-hand side of (\ref{eq:Phidot}).
Namely, {\bf case (ii)}:
\begin{subequations}\label{eq:case2}
\begin{align}
\psi - \alpha \big( \shift^m \partial_m \psi
- N\Pi\big) &  = B_\psi
\label{eq:psiimp2}\\
\Pi - \alpha \big(
\shift^m \partial_m \Pi
- N g^{jm} \partial_j \Phi_m + \epsilon NS\big)
&  = B_\Pi
\label{eq:Piimp2}\\
\Phi_k - \alpha \big( \shift^m \partial_m \Phi_k
- N \partial_k \Pi + \Phi_m \partial_k \shift^m - \Pi \partial_k N\big)
&  = B_{\Phi_{k}}.
\label{eq:Phiimp2}
\end{align}
\end{subequations}
Finally, treating all or nearly all terms implicitly, we arrive at 
{\bf case (iii)}:
\begin{subequations}\label{eq:case3}
\begin{align}
\psi - \alpha \big( \shift^m \partial_m \psi
- N\Pi\big) &  = B_\psi
\label{eq:psiimp3}\\
\Pi - \alpha \big(
\shift^m \partial_m \Pi
- N g^{jm} \partial_j \Phi_m
+ N K \Pi
+ N \Phi_m J^m
+ \epsilon N S\big)
&  = B_\Pi
\label{eq:Piimp3}\\
\Phi_k - \alpha \big( \shift^m \partial_m \Phi_k
- N \partial_k \Pi + \Phi_m \partial_k \shift^m - \Pi \partial_k N\big)
&  = B_{\Phi_{k}}.
\label{eq:Phiimp3}
\end{align}
\end{subequations}
For each of our three cases, we note that the inhomogeneity 
$\boldsymbol{B} = \{B_\psi,B_\Pi,B_{\Phi_k}\}$
corresponds to the term in (\ref{eq:abstractimpeqn}) built with ARK 
stage values, cf.~Eq.~(\ref{eq:uidef}). While we have considered 
only three possible IMEX splittings (really six including the 
$\epsilon = 0,1$ choice), other variations are of course possible.     
Ignoring the subcases afforded by the choice of
$\epsilon$, case (i) corresponds to the minimal implicit sector 
for which our methods are applicable, case (iii) to the fully 
implicit scenario, and case (ii) to a scenario in the middle.
Note that for cases (ii) and (iii) the gradient of the 
left-hand side of the $\psi$ equation [Eqs.~(\ref{eq:psiimp2}) and
  (\ref{eq:psiimp3}), respectively] gives the corresponding
  $\Phi_k$ equation [Eqs.~(\ref{eq:Phiimp2}) and (\ref{eq:Phiimp3}),
  respectively], up to the replacement $\partial_k\psi\to\Phi_k$. 
 This mirrors the structure of the first-order PDE,
  cf.~the remark after Eq.~(\ref{eq:dt_C}).

While we do not solve these first-order systems numerically, 
we expect that the following theoretical considerations are relevant.
We view each set [Eqs.~(\ref{eq:case1}), (\ref{eq:case2}), 
or (\ref{eq:case3})] 
of implicit equations as a {\em spatial} boundary value problem subject
to Dirichlet boundary conditions on the same characteristic fields
(\ref{eq:charvars}) as those described in the last paragraph of
\ref{subsec:FirstOrderScalarWave}. In other words, the choice of
boundary data for these implicit solves corresponds to the same
boundary data controlled in the {\em evolution}
initial-boundary-value problem. This physically reasonable viewpoint
is analyzed further in an appendix. On the outer boundary $\Bouter$
where $\shift^k n_k > 0$, we fix $Z^1$, $Z^2_j$, and $U^{1-}$ as
boundary data.  Typically, $V^kn_k<0$ on $\Binner$, 
so no boundary conditions on $Z^1$ and $Z^2_j$ are imposed.  
If $-V^kn_k-N<0$ on $\Binner$, then we would fix $U^{1-}$ as 
boundary data. This inequality would hold, for example, in 
our Schwarzschild scenario,
provided $\Binner$ were chosen as a surface {\em outside} the horizon.
Now, each set [Eqs.~(\ref{eq:case1}), (\ref{eq:case2}), 
or (\ref{eq:case3})]
involves the first-order derivatives of 5 fields, whence we expect
that 5 boundary conditions are needed to uniquely determine a
solution. Indeed, $\Pi$ and $\Phi_k$ should be determined by
(\ref{eq:case1}b,c), (\ref{eq:case2}b,c) or (\ref{eq:case3}b,c) and
specification of the following 4 boundary conditions: $Z^2_j$ and
$U^{1-}$ on $\Bouter$, and $U^{1-}$ on $\Binner$ (provided $-n_k
\shift^k - N < 0$). Once $\Pi$ is known, the remaining equation for
$\psi$ could then be integrated subject to a remaining fifth boundary
condition for $Z^1$ on $\Bouter$.  We analyze a simplified system which
justifies this counting argument in Appendix~\ref{sec:BCSfirstImpEQNS}.

For the Schwarzschild scenario when $\Binner$ lies
inside the horizon, the situation is different.  Let us view
$\Binner$ as spherically symmetric, and let us extend the normal
$n^k$ to $\Binner$ smoothly into the volume such that $n^k$ is normal to
$r=\mbox{const}$ spheres, and normalized such that $g_{ij}n^in^j=1$. 
Combination of the first-order implicit equations for 
$\Pi$ and $\Phi_k$ yields 
\begin{equation}\label{eq:U1-Implicit}
U^{1-} - \alpha \left[(\shift^k + N n^k) \partial_k U^{1-} + \cdots
\right] = B_\Pi - n^k 
B_{\Phi_k},
\end{equation}
where we define $U^{1-}=\Pi-n^k\Phi_k$ even away from $\Binner$.  On the
horizon $V^k+N n^k=0$, and thus Eq.~(\ref{eq:U1-Implicit}) determines
$U^{1-}$ algebraically.  Integration of Eq.~(\ref{eq:U1-Implicit}) inward from
the horizon to $\Binner$ then results in the value of $U^{1-}$
on $\Binner$.  Thus $U^{1-}$ on $\Binner$ is determined 
self-consistently by the equations, and we are not free to pick it.
  
As with the evolution initial-boundary-value problem, in our implicit
boundary value problems we relate some boundary data to the constraint
$\mathcal{C}_k$. First, we identify the tangential components 
$P^k_j \mathcal{C}_k|_\Bouter$ with the boundary data 
$Z^2_j |_\Bouter$. In other words, on $\Bouter$ we set 
$Z^2_j = P^k_j (\partial_k\psi - \mathcal{C}_k)$, where 
$P^k_j\mathcal{C}_k$ is a fixed function (typically zero). 
Along with the boundary data
$U^{1-}$, these tangential components then allow for
recovery of $\Pi$ and $\Phi_k$. Second, writing 
(\ref{eq:psiimp1}), (\ref{eq:psiimp2}), or (\ref{eq:psiimp3}) as
\begin{equation}
\psi - \alpha \shift^k \mathcal{C}_k = \alpha (\shift^k\Phi_k - N\Pi) 
+ B_\psi,
\end{equation}
we may now view all terms on the right-hand side as a given source.
Rather than fixing $Z^1 = \psi$ as boundary data on $\Bouter$, we 
equivalently fix
$\shift^k\mathcal{C}_k|_\Bouter$, since $\psi|_\Bouter$ can
then be recovered from the last expression evaluated at $\Bouter$.
We may then formally view our outer boundary conditions
on $\Bouter$ as controlling $U^{1-}$ and $\mathcal{C}_k$.

For cases (ii) and (iii) the listed implicit equations determine an 
implicit equation for the constraint.
Indeed, notice that the pairs (\ref{eq:psiimp2}),(\ref{eq:Phiimp2}) and
(\ref{eq:psiimp3}),(\ref{eq:Phiimp3}) are the same. If we subtract, say,
(\ref{eq:Phiimp3}) from the Cartesian derivative of (\ref{eq:psiimp3}),
then we arrive at
\begin{equation}
\mathcal{C}_k - \alpha \big( \shift^m \partial_m \mathcal{C}_k
+ \mathcal{C}_m \partial_k \shift^m\big)
= \partial_k B_\psi - B_{\Phi_{k}},
\label{eq:constraintimp}
\end{equation}
an equation we may alternatively express in terms of the Lie derivative as
\begin{equation}
\mathcal{C}_k - \alpha \pounds_V \mathcal{C}_k
= \partial_k B_\psi - B_{\Phi_{k}}.
\label{eq:lieconstraintimp}
\end{equation}
Contraction of (\ref{eq:constraintimp}) on $\alpha\shift^k$ yields
\begin{equation}
\alpha \shift^k\mathcal{C}_k - \alpha^2 \shift^j \partial_j
(\shift^k\mathcal{C}_k)
= \alpha\shift^k (\partial_k B_\psi - B_{\Phi_k}).
\label{eq:shiftconstraintimp}
\end{equation}
In principle, Eq.~(\ref{eq:lieconstraintimp}) might be integrated 
along the shift, say inward from the outer boundary $\Bouter$ 
where Dirichlet boundary conditions on $\mathcal{C}_k$ are set. 

The ARK scheme should preserve the 
constraint; i.e.~if $\mathcal{C}_k=0$ initially, it should remain zero.   
We investigate this point by
combining  (\ref{eq:uidef}) and (\ref{eq:abstractimpeqn}) into
a formula for the $i$th stage source,
\begin{align}
\boldsymbol{B}^{(i)}
= \boldsymbol{u}^{n}
+ \Delta t \sum_{j=1}^{i-1} \left[
a^E_{ij} \boldsymbol{f}^E(t^{(j)},\boldsymbol{u}^{(j)})
+ a^I_{ij} \boldsymbol{f}^I(t^{(j)}, \boldsymbol{u}^{(j)})\right]
,\quad 2 \le i \leq s,
\label{eq:Bexpression}
\end{align}
where we recall that $\boldsymbol{u}^{(1)} = \boldsymbol{u}^{n}$.
For both cases (ii) and (iii), the $\psi$ and $\Phi_k$ components 
of $\boldsymbol{f}^E(t,\boldsymbol{u})$ vanish, whereas
\begin{equation}
\boldsymbol{f}^I(t,\boldsymbol{u})
= \left(\begin{array}{c}
\shift^k \partial_k \psi - N\Pi\\
\bullet\\
 \shift^m \partial_m \Phi_k
- N \partial_k \Pi + \Phi_m \partial_k \shift^m - \Pi \partial_k N
\end{array}
\right),
\end{equation}
with $\bullet$ indicating an expression irrelevant for the present 
discussion. 
Therefore, if the previous stage values $\boldsymbol{u}^{(j)}$, 
$j = 1,\ldots,i-1,$ obey the constraint (\ref{eq:C}), that 
is $\mathcal{C}^{(1)}_k = \cdots = \mathcal{C}^{(i-1)}_k = 0$, 
then the $i$th source $\boldsymbol{B}^{(i)}$ will satisfy 
$\partial_k B^{(i)}_\psi = B^{(i)}_{\Phi_k}$. As a result, 
(\ref{eq:constraintimp}) will be a homogeneous equation for 
the constraint $\mathcal{C}^{(i)}_k$  at the $i$th stage,
with the solution $\mathcal{C}_k^{(i)}=0$ in the interior because
$C_k^{(i)}=0$ has been enforced on the boundary.
We will draw on these observations below.

\subsection{Second-order implicit equation}

In principle, one could solve directly the first-order implicit
equations given in Eqs.~(\ref{eq:case1}), (\ref{eq:case2}),
or~(\ref{eq:case3}), and we have done so in spherical symmetry.  
However, for the more demanding 3d cases leading toward our ultimate goal 
of handling binary black holes, we would like to use the 
multidomain spectral 
{\tt EllipticSolver}~\cite{Pfeiffer2003} which is part of the 
Spectral Einstein Code 
{\tt SpEC} used for binary black hole evolutions~\cite{Kidder2000a,Scheel2006,Boyle2007}. 
The {\tt EllipticSolver} has
been written to handle second-order elliptic equations. Moreover,
preconditioning strategies for second-order elliptic equations are
well understood relative to those for first-order equations.  For
these reasons, we have chosen not to directly solve first-order
equations. Rather, we first solve a single second-order scalar
equation for $\psi$, one stemming from combination of the above
equations and subject to appropriate boundary conditions discussed
below. This $\psi$ equation is different for each of the three cases,
and it is only for cases (ii) and (iii) that we can show, at least
formally, that our solution process is consistent with solving the
original first-order set of equations. Once a solution $\psi$ has
been determined for each case, we obtain $\Pi$ algebraically using
(\ref{eq:psiimp1}), (\ref{eq:psiimp2}), or 
(\ref{eq:psiimp3}), all the same equation. 
Finally, we recover $\Phi_k$ from $\psi$ via
differentiation. Therefore, at each stage in our IMEX algorithms we
perform what amounts to a naive constraint projection. This is 
necessary for case (i), but would seem not strictly necessary for 
cases (ii) and (iii). 

Both ARK3 and ARK4 have explicit first stages, for which no implicit
solve needs to be done.  Nevertheless, in order to achieve stability
in our IMEX evolutions, we must perform the naive constraint 
projection on the first-stage fields, at least when such projection 
is carried out on the other stages. The discussion after 
Eq.~(\ref{eq:Bexpression}) pertains to exact arithmetic, whereas 
round off errors in the stage expansion (\ref{eq:unp1}) will result 
in a $\boldsymbol{u}^{n+1}$ which violates the constraints. While
this violation is negligible over a single time-step, such
violations appear to accumulate. Projection of the 
first-stage fields guarantees that $\partial_k B^{(1)}_\psi = 
B^{(1)}_{\Phi_k}$ throughout the evolution.

Let us first derive the scalar equation for $\psi$ associated 
with case (i), Eqs.~(\ref{eq:case1}). Combination of 
(\ref{eq:case1}a,b) eliminates the term 
proportional to $\Pi$,
\begin{align}
\psi - \alpha  \shift^m\partial_m\psi + \alpha^2 N \shift^m
\partial_m\Pi - & \alpha^2 N^2 g^{jm} \partial_j\Phi_m
= B_\psi - \alpha N B_\Pi - \epsilon\alpha^2 N^2 S,
\label{eq:psiPicombo}
\end{align}
whereas from (\ref{eq:Phiimp1}) we obtain
\begin{equation}
\alpha \shift^m\Phi_m - \alpha^2  \shift^j \shift^m \partial_j\Phi_m
+ \alpha^2 N \shift^m \partial_m\Pi
= \alpha \shift^m B_{\Phi_m}.
\end{equation}
The difference of the last two equations is independent of $\Pi$,
\begin{align}
\psi - \alpha  \shift^m (\partial_m\psi + \Phi_m)
- \alpha^2  \left( N^2 g^{jm} - \shift^j \shift^m\right)
\partial_j\Phi_m =
B_\psi - \alpha N B_\Pi - \alpha  \shift^m B_{\Phi_m}
- \epsilon\alpha^2 N^2 S.
\label{eq:psielliptic1a}
\end{align}
Next, we use the constraint~(\ref{eq:C}) 
to replace $\Phi_k$ by $\partial_k\psi-{\cal C}_k$, and find 
\begin{align}
\psi - & 2\alpha  \shift^j \partial_j\psi
- \alpha^2  \left( N^2 g^{jk} - \shift^j \shift^k\right)
\partial_j\partial_k \psi = 
\nonumber \\
& 
B_\psi - \alpha N B_\Pi - \alpha  \shift^k B_{\Phi_k}
- \epsilon\alpha^2 N^2 S
-  \alpha \shift^j\mathcal{C}_j 
- \alpha^2 (N^2 g^{jk} - \shift^j \shift^k)\partial_j \mathcal{C}_k,
\label{eq:smallelliptic}
\end{align}
which is our second-order $\psi$ equation for case (i).

Derivation of the $\psi$ equation for cases (ii) and (iii) is 
more complicated, but nevertheless follows the same steps taken for 
case (i). For example, Eqs.~(\ref{eq:case2}a,b) again lead to 
(\ref{eq:psiPicombo}). Similar to before, we eliminate the 
term $\alpha^2 N \shift^m \partial_m \Pi$ with the contraction of
$\alpha V^k$ on the $\Phi_k$ equation (\ref{eq:Phiimp2}). However, 
now this third equation is more complicated and features an extra 
factor of $\Pi$. Therefore,
we first use (\ref{eq:psiimp2}) to rewrite (\ref{eq:Phiimp2}) as
\begin{equation}
\Phi_k - a_k \psi - \alpha \big(
\shift^m \partial_m \Phi_k
- N \partial_k \Pi + \Phi_m \partial_k \shift^m 
- a_k V^m \partial_m \psi\big) = B_{\Phi_{k}} - a_k B_\psi,
\end{equation}
where $a_k \equiv \partial_k \log N$. Finally, we contract the 
last equation on $\alpha V^k$, subtract the result from 
(\ref{eq:psiPicombo}), and then use (\ref{eq:C}) to replace 
all $\Phi_k$ terms by $\partial_k\psi - \mathcal{C}_k$. These
steps yield the following equation for case (ii):
\begin{align}
\big(1+\alpha \shift^k a_k\big)\psi
& -\big[2\alpha \shift^k
  +\alpha^2 (a_j \shift^j \shift^k 
- \shift^j \partial_j \shift^k)\big]\partial_k\psi
-\alpha^2 \big(N^2 g^{jk} - \shift^j \shift^k\big)
\partial_j\partial_k\psi
\nonumber \\
&
 = (1+\alpha \shift^k a_k) B_\psi - \alpha N B_\Pi - \alpha \shift^k 
B_{\Phi_k} - \epsilon\alpha^2 N^2 S
\nonumber \\
&\quad - \alpha^2 N^2 g^{jk}\partial_j \mathcal{C}_k
+ \alpha^2 V^j\partial_j (V^k\mathcal{C}_k)
-\alpha V^k\mathcal{C}_k.
\label{eq:mediumelliptic}
\end{align}
Even more involved calculations using Eqs.~(\ref{eq:case3})
similarly yield
\begin{align}
\big[1 + \alpha\big(\shift^k a_k
- NK\big)\big]\psi
- \big[ & 2\alpha\shift^k
+\alpha^2\big(
a_j \shift^j \shift^k
- \shift^j \partial_j \shift^k
- N^2 J^k
-\shift^k NK
\big)
\big]\partial_k \psi
\nonumber \\
-\alpha^2\big(N^2 g^{jk}-\shift^j \shift^k\big)\partial_j\partial_k\psi
& =
\big[1+\alpha
\big(\shift^k a_k - NK\big)\big]B_\psi - \alpha N B_\Pi
- \alpha
\shift^k B_{\Phi_k} 
\nonumber \\
&\quad - \epsilon\alpha^2 N^2 S
+\alpha^2 N^2 \big(
J^k \mathcal{C}_k
-g^{jk}\partial_j \mathcal{C}_k\big)
\nonumber \\
&\quad + \alpha^2 V^j\partial_j (V^k\mathcal{C}_k)
-\alpha V^k\mathcal{C}_k,
\label{eq:largeelliptic}
\end{align}
the second-order $\psi$ equation for case (iii). Notice that neither
(\ref{eq:smallelliptic}), (\ref{eq:mediumelliptic}), nor
(\ref{eq:largeelliptic}) features derivatives of $\boldsymbol{B} =
(B_\psi, B_\Pi, B_{\Phi_k})$. The absence of $\boldsymbol{B}$
derivatives indicates that we have not differentiated any of our
original first-order equations.

Let us now consider the corresponding boundary
conditions for the second-order equations (\ref{eq:smallelliptic}),
(\ref{eq:mediumelliptic}), (\ref{eq:largeelliptic}). Combination of
(\ref{eq:psiimp1}), the same for all cases, with the formula 
(\ref{eq:charvars}) for $U^{1-}$ yields
\begin{equation}
\psi - \alpha \shift^m \partial_m \psi
+\alpha N\big(U^{1-}+n^k \Phi_k\big) = B_\psi.
\end{equation}
Using the constraint $\mathcal{C}_k = \partial_k\psi - \Phi_k$, we
therefore find
\begin{equation}
\psi
+ \alpha \big(N n^m - \shift^m\big) \partial_m \psi
 = B_\psi - \alpha N U^{1-} + \alpha N n^k \mathcal{C}_k
\label{eq:psiBCUmC}
\end{equation}
as our boundary condition.  If $\Binner$ lies inside the horizon, we
shall not impose (\ref{eq:psiBCUmC}) on $\Binner$, as the initial
boundary value problem for Eqs.~(\ref{eq:FirstOrderScalarWave}) does not
require an inner boundary condition.  For a Schwarzschild spacetime in
Kerr-Schild coordinates, Sec.~\ref{sec:Schwarzschild} shows that an
outer boundary condition alone is indeed sufficient.  In this setting,
the inner boundary condition for the second-order $\psi$ equation is
replaced by a regularity condition across the horizon.  Numerical
tests in Sec.~\ref{sec:numtests} indicate that this viewpoint holds in
a more general setting.
While we have been careful to retain all constraint 
terms in deriving 
Eqs.~(\ref{eq:smallelliptic}),~(\ref{eq:mediumelliptic}),~(\ref{eq:largeelliptic})
and (\ref{eq:psiBCUmC}), 
in practice we have set $\mathcal{C}_k=0$ before solving 
these equations numerically.

At least for cases (ii) and (iii), the following argument formally 
proves that ---in lieu of directly solving the full set of 
first-order implicit equations (\ref{eq:case2}) or 
(\ref{eq:case3})--- we may instead take the following steps:
solve (\ref{eq:lieconstraintimp}), solve either 
(\ref{eq:mediumelliptic}) or (\ref{eq:largeelliptic}) (the 
single second-order $\psi$ equation), and 
then recover $\Pi$ from Eq.~(\ref{eq:psiimp2}) or 
(\ref{eq:psiimp3})
and $\Phi_k$ from Eq.~(\ref{eq:C}). In combining the 
first-order equations to produce the single second-order 
 $\psi$ equation, we have 
not differentiated the original equations, rather second 
derivative terms have arisen via substitution with the 
constraint $\partial_j\Phi_k \rightarrow \partial_j\partial_k\psi 
- \partial_j\mathcal{C}_k$. 
Therefore, the $\psi$ equation is truly 
a second-order equation only if we interpret the constraint terms 
appearing on the right-hand side as part of the inhomogeneity. To 
achieve this interpretation, we assume that we may integrate the 
system (\ref{eq:lieconstraintimp}), inward from the outer boundary 
$\Bouter$, where we fix $\mathcal{C}_k |_\Bouter$. These
boundary conditions are consistent, as we have earlier argued that
$\mathcal{C}_k |_\Bouter$ is part of the boundary data for the 
original first-order set of implicit equations.

\subsection{Schwarzschild geometry}
\label{sec:Schwarzschild}

We now specialize the above equations to the Schwarzschild 
geometry written in Kerr-Schild coordinates. The line-element is
\begin{equation}
\mathrm{d}s^2 = -N^2\mathrm{d}t^2 + L^2\big(\mathrm{d}r +  \shift^r
\mathrm{d}t\big)^2 + r^2\big(\mathrm{d}\theta^2 
+\sin^2\theta\mathrm{d}\phi^2\big),
\label{eq:KerrSchild}
\end{equation}
where the lapse, radial lapse, and shift are given in terms
of the mass parameter $M$ by
\begin{equation}
N = \sqrt{\frac{r}{r+2M}},\quad
L = \sqrt{\frac{r+2M}{r}},\quad
\shift^r = \frac{2M}{r+2M}.
\label{eq:geometry}
\end{equation}
This is the same line-element as given in Eq.~(59) of 
\cite{Holst2004}. 

Consider the Cartesian coordinates $x^k$ stemming from the polar
coordinates $(r,\theta,\phi)$ via the standard formulas:
$(x,y,z) = (r\sin\theta\cos\phi,r\sin\theta\sin\phi,r\cos\theta)$. 
We introduce a radial vector $\nu^k = x^k/r$ which is not normalized 
with respect to the spatial metric determined by 
(\ref{eq:KerrSchild},\ref{eq:geometry}). 
The vector $n^k = L^{-1}\nu^k$ is the outward-pointing unit normal to 
the spherical foliation of a spacelike level-$t$ hypersurface.
With respect to the Cartesian coordinates $x^k$, we may express the 
spatial metric and inverse metric as follows:
\begin{equation}
g_{jk} = (L^2 -1) (\partial_j r)(\partial_k r) + \delta_{jk},
\qquad
g^{jk} = (L^{-2} -1)\nu^j \nu^k + \delta^{jk}.
\label{eq:3metric}
\end{equation}
Here $\delta_{jk} = \delta^{jk} = \mathrm{diag}(1,1,1)$ is the 
flat metric, and $\partial_j r = \delta_{jk}\nu^k$. To 
avoid ambiguity
as to which metric ($g_{jk}$ or $\delta_{jk}$) has been used to lower
the index, we will not write $\nu_k$. With the above 
formulas for the Cartesian components of the metric and
Eqs.~(\ref{eq:Jk},\ref{eq:K}), we find that 
\begin{equation}
J^k =
\frac{1}{L^2}
\left[\frac{L'}{L} -\frac{N'}{N} 
+\frac{2}{r}(L^2-1)\right]\nu^k,
\qquad
K = \frac{1}{N}\left[(\shift^r)'+ \frac{2}{r}\shift^r 
+ \frac{L'}{L}\shift^r
\right].
\label{JkandKforSch}
\end{equation}
To reach these equations, we have used the following identities 
(valid in the Cartesian coordinate system): $g^{1/2} = L$, 
$V^k = V^r \nu^k$, $\nu^j \partial_j \nu^k = 0$, and $\partial_j\nu^k = 
r^{-1} (\delta^k_j - \nu_j \nu^k)$. 
Equations~(\ref{eq:3metric}) and~(\ref{JkandKforSch}) 
hold for any choice of $N$, $L$ and $\shift^r$.  Specializing to 
the values given in Eq.~(\ref{eq:geometry}), we obtain
\begin{equation}
g_{jk}=\delta_{jk}+\frac{2M}{r}(\partial_j r)(\partial_k r),
\qquad
J^k = \frac{2M(r+4M)}{r(r+2M)^2}
\nu^k,\qquad
N K = \frac{2M(r+3M)}{r(r+2M)^2}.
\label{eq:JkandEnnKay}
\end{equation}

For the chosen Schwarzschild background and coordinates, 
we now show that our solution procedure 
involving the second-order $\psi$ equation 
(\ref{eq:mediumelliptic}) or (\ref{eq:largeelliptic})
is equivalent to solving the original set of
first-order equations (\ref{eq:case2}) or (\ref{eq:case3}).  
In establishing 
this claim we must show that (\ref{eq:constraintimp}) can be 
integrated inward from the outer boundary $\Bouter$ 
at $r = r_\mathrm{max}$, and that a 
{\em regular} solution to the $\psi$ equation is determined 
by the outer boundary condition alone. We consider the integration 
of (\ref{eq:constraintimp}) in Appendix \ref{sec:impC}, and turn 
to the latter issue now.  Whereas integration 
of (\ref{eq:constraintimp}) is only relevant for cases (ii) 
and (iii), the issue of a regular solution to the $\psi$ equation 
also pertains to (\ref{eq:smallelliptic}), and so we include this
equation in our analysis.  Each of the second-order 
scalar equations [Eq.~(\ref{eq:smallelliptic}), (\ref{eq:mediumelliptic}),
or (\ref{eq:largeelliptic})]
takes the following form:
\begin{equation}
\mathcal{R}(r) \psi + \alpha\mathcal{S}(r) \nu^k \partial_k\psi
+\alpha^2\left(N^2 g^{jk}-\shift^j \shift^k\right)\partial_j\partial_k\psi
=\mathcal{G}.
\label{eq:generalelliptic}
\end{equation}
Here we view constraint terms appearing 
in $\mathcal{G}$, if present, as predetermined via 
integration of (\ref{eq:constraintimp}).  We continue by calculating
\begin{equation}
g^{jk}\partial_j\partial_k\psi
= L^{-2}\partial_r^2\psi + 2r^{-1}\partial_r\psi 
+  r^{-2} \Delta_{S^2}\psi,
\end{equation}
where $\Delta_{S^2}$ is the
Laplace operator associated with $S^2$, the unit-radius round sphere.  
Next, we set $\psi = \psi_{\ell m}(r) 
Y_{\ell m} (\theta,\phi)$ in (\ref{eq:generalelliptic}), thereby 
obtaining the equation for a generic spherical-harmonic mode,
\begin{equation}
\left[\mathcal{R}(r)
-\frac{\alpha^2 N^2\ell(\ell +1)}{r^2}\right]
\psi_{\ell m} + 
\left[\alpha\mathcal{S}(r)
+\frac{2\alpha^2 N^2}{r}\right]\partial_r\psi_{\ell m}
+\alpha^2\left[\frac{N^2}{L^{2}}-(\shift^r)^2\right]
\partial^2_r\psi_{\ell m}
= \mathcal{G}_{\ell m}.
\end{equation}
This equation has the form
\begin{equation}\label{eq:appendixeqn}
\mathcal{Q}(r)
w + \alpha\mathcal{P}(r)w'
+\alpha^2(r-2M)w''
= h(r),
\end{equation}
with
\begin{equation}
\mathcal{Q}(r) = (r + 2M)\left[\mathcal{R}(r)
-\frac{\alpha^2 N^2\ell(\ell +1)}{r^2}\right],
\quad \mathcal{P}(r) =  (r + 2M)\left[\mathcal{S}(r)+\frac{2\alpha 
N^2}{r}\right].
\end{equation}
Note that the coefficient of the second-order term $w''$ passes through 
zero at the black hole horizon, $r=2M$.  We study 
Eq.~(\ref{eq:appendixeqn}) in Appendix \ref{sec:SBVP}, 
where we show that $r=2M$ is a regular singular 
point. Furthermore, in the appendix we compute 
the indicial exponents associated with the singular point,
and argue that, despite the second-order character of
(\ref{eq:appendixeqn}), an outer boundary condition alone
determines a unique solution which is regular up to and even 
across the horizon.

Thus the following picture emerges: If the radius $r_{\rm min}$ of
$\Binner$ satisfies $r_{\rm min}>2M$, then the scalar wave equation 
requires a boundary condition for $U^{1-}$ on {\em both} $\Binner$ 
and $\Bouter$; the
corresponding second-order implicit equation
(\ref{eq:generalelliptic}) is everywhere regular and requires boundary
conditions on {\em both} boundaries as well.  For $r_{\rm min}<2M$,
the inner boundary is an outflow boundary for the scalar wave
equation, and a boundary condition on  $U^{1-}$ is necessary only on 
$\Bouter$; in this case, a unique solution to
Eq.~(\ref{eq:generalelliptic}) is determined by an outer boundary
condition alone, along with the assumption that the solution is regular 
across the horizon.

\section{Numerical tests}\label{sec:numtests}

\subsection{Comparison with explicit time-stepping}\label{subsec:compare}
In this first subsection we demonstrate that our numerical 
IMEX algorithm can solve a standard initial value problem. 
To do so, we consider the evolution of pulse initial data, 
a problem for which an explicit algorithm would be better 
suited. Here we are evolving the wave equation 
$\nabla_\mu\nabla^\mu\psi=0$ without a source term,
that is for $S=0$.

\begin{figure}[t]
\begin{center}
\includegraphics[width=12.0cm,bb=-44 140 656 690]{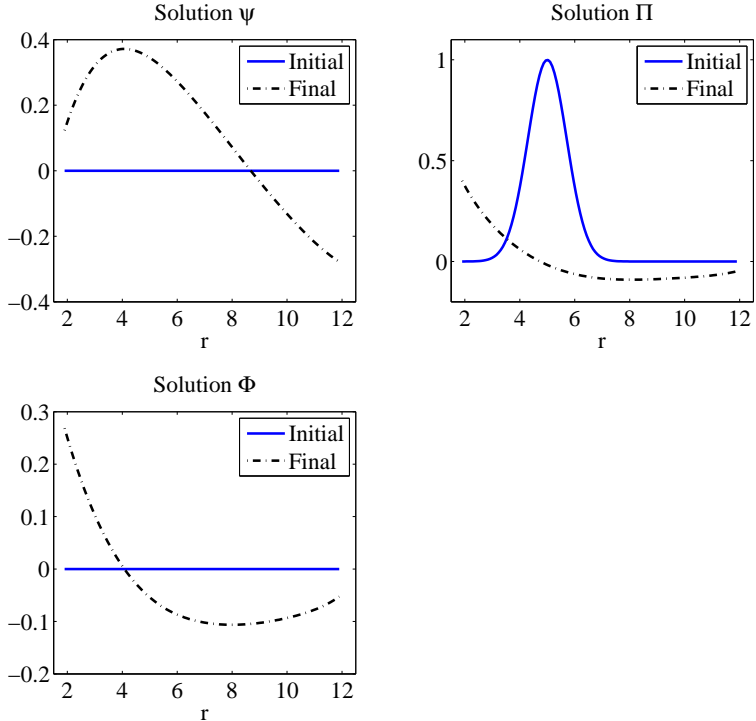}
\end{center}
\caption{Field configurations at initial and final times.}
\label{fig:ERKrun}
\end{figure}
\begin{figure}[t]
\begin{center}
\includegraphics[width=12.0cm,bb=-44 130 656 690]{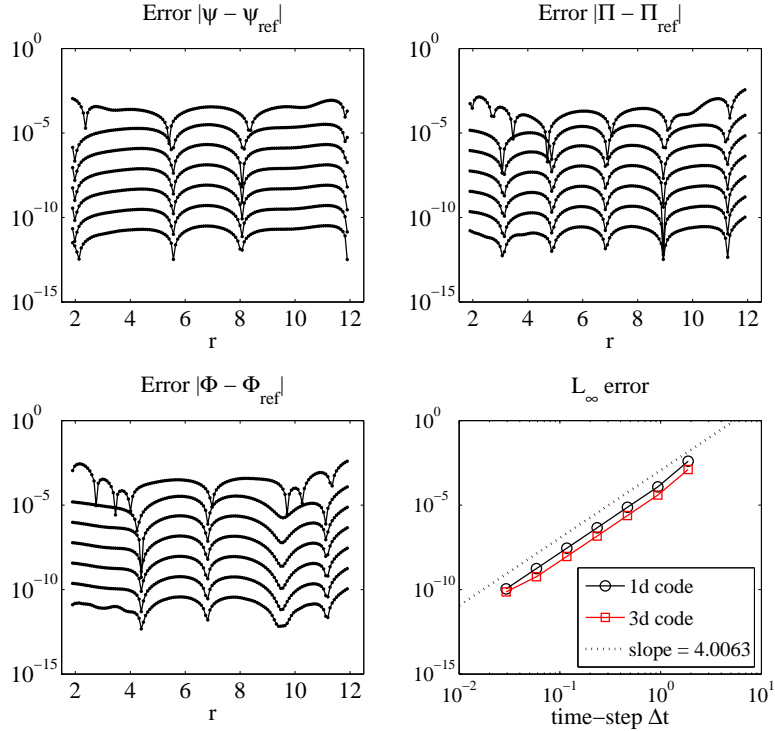}
\end{center}
\caption{Error of implicit evolutions relative to the explicit
reference solution. The dotted line is a least-squares fit of
the last five 1d data points, although shifted to make for a
better figure.}
\label{fig:ARK4}
\end{figure}

The following experiment has been carried out both with a 
one-dimensional radial code (in {\sc Matlab}) and with a 
three-dimensional code  (in {\tt SpEC}). Whereas the 3d
code uses the variables $\{\psi,\Pi,\Phi_x,\Phi_y,\Phi_z\}$,
the 1d code uses $\{\psi,\Pi,\Phi = \partial_r\psi\}$. On 
the radial domain $[1.9,11.9]$, initial data
\begin{align}
\text{1d: }\quad \psi&=0,& \Pi &= \exp\left[-(r-5)^2\right],&
\Phi\;\,&=0,
\nonumber \\
\label{eq:GaussianPacket}
\qquad\qquad\text{3d: }\quad \psi&=0,
& \Pi &=  
\exp\left[-(r-5)^2\right]
\text{Re}\big[Y_{11}(\theta,\phi)\big],&
 \Phi_k&=0,\qquad\qquad
\end{align}
is evolved to time $t_\mathrm{final} = 15$, with the background 
geometry taken as $M = 1$ Schwarzschild in Kerr-Schild 
coordinates. Initial and final radial mode profiles are depicted 
in Fig.~\ref{fig:ERKrun}. The first step of the experiment is 
to generate a reference solution, using an explicit Runge-Kutta 
(ERK) time-stepper, either (for 1d) the classical fourth-order 
scheme or (for 3d) the fifth-order Cash-Karp scheme
\cite{CashKarp1990}. In both cases we choose a fixed time-step 
$\Delta t \simeq 0.00366$, and so are not using the potential 
adaptivity of the Cash-Karp scheme. For both the 1d and 3d 
experiments, we place no boundary condition at the inner 
radius $r = 1.9$, and a Sommerfeld boundary condition 
$U^{1-}=0$
[cf.~Eq.~(\ref{eq:charvars}), either 
$\Pi - \Phi = 0$ in 1d or $\Pi - n^k\Phi_k = 0$ in 3d] at the 
outer boundary $r = 11.9$.  
We further enforce constraint-preserving outer boundary 
conditions which are analogous to the boundary 
conditions applied to the black hole evolutions 
in~\cite{Lindblom2006}.  For the scalar characteristic 
field $Z^1$ and the 3d code, we use 
\begin{equation}
\partial_t Z^1 = -N\Pi + \shift^k\Phi_k,
\end{equation}
cf.~Eq.~(40) Ref.~\cite{Holst2004}.  For the 1d code
we similarly use $\partial_t Z^1 = -N\Pi + \shift^r\Phi$
as the outer boundary condition.
For $Z^2_i$, we employ the analogue of Eq.~(65) 
in Ref.\cite{Lindblom2006},
\begin{equation}
\partial_t Z^2_i = D_t Z^2_i 
- n_k \shift^k n^m
\big(\partial_m \Phi_i - \partial_i \Phi_m\big).
\end{equation}
Here $D_t Z^2_i$ denotes the $P^j_i$ projection of the 
right-hand side of Eq.~(\ref{eq:Phidot}). We use one spherical 
shell, with $61$ radial collocation points. We fix the angular 
resolution for the 3d evolution with $\ell_\mathrm{max} = 5$ as 
the top spherical-harmonic index.  
The explicit evolution uses the same angular filtering
  as~\cite{Holst2004}.  When the right-hand sides of
Eqs.~(\ref{eq:FirstOrderScalarWave}) are computed, they are
transformed to scalar spherical harmonics (for the $\psi$ and $\Pi$
components) or vector spherical harmonics (for the
$\Phi_k$ component), and the top two modes are truncated.  
\begin{figure}[t]
\begin{center}
\includegraphics[width=11.0cm]{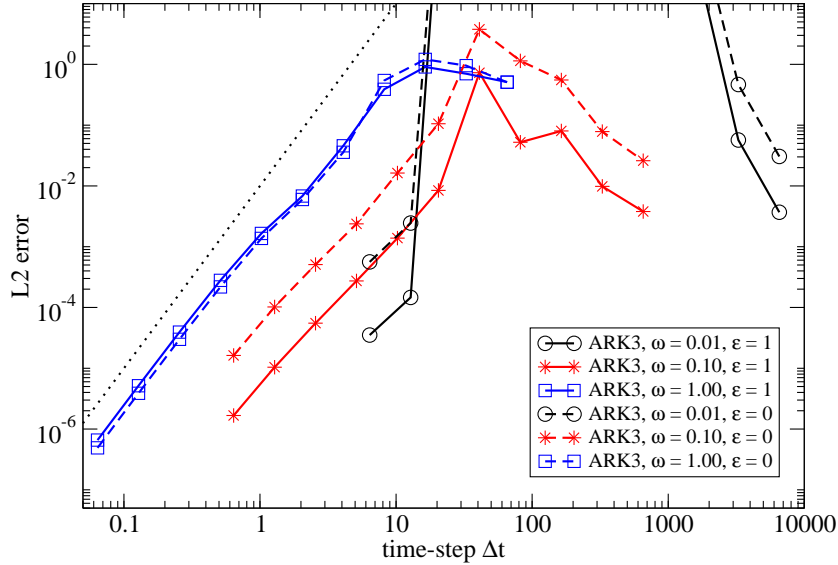}
\end{center}
\caption{Performance of ARK3 for case (i) IMEX splitting.
The dotted line corresponds to exact third-order convergence.}
\label{fig:ARK3case_i}
\end{figure}

We next carry out the same evolution via IMEX evolutions taking
case (iii) (the $\epsilon$ value is irrelevant because $S=0$),
and check the results against the reference solution.
During an evolution, the requisite implicit solves have been carried
out with the {\tt EllipticSolver} in {\tt SpEC}, as described in
\cite{Pfeiffer2003}. For this simple class of problems,
we have chosen finite-difference preconditioning. Precisely,
we have used an exact LU decomposition of a finite-difference
approximation $A_{FD}$ to the operator associated with the second-order
$\psi$ equation. The {\tt EllipticSolver} interfaces with {\tt petsc}'s
iterative solvers, and for the iterative linear solves we have
used GMRES, choosing all error tolerances close to machine precision.
As mentioned, for these experiments $r_\mathrm{min}=1.9M<2M$, 
so the inner boundary lies inside of the horizon. Therefore, 
throughout our evolutions we solve the second-order $\psi$ equation 
with no boundary condition at the innermost collocation points. Rather
in the relevant matrix-vector multiply needed for the iterative
solver, only the PDE is enforced at these points.
The elliptic equation for $\psi$ is solved for $Y_{\ell m}$ modes
  with $\ell\leq \ell_{\rm max}-2$ to avoid technical issues due to
  representation of Cartesian tensor components 
  with scalar spherical harmonics.  This acts as an angular filter for
  $\psi$, obviating the need for 
  further angular filtering as described above for the explicit evolution.

For a number of temporal resolutions and for 
the ARK4 method, we show the results in Fig.~\ref{fig:ARK4}. The
errors plotted in the first, second, and third quadrants correspond
to the 1d code, and these errors have been computed after 
interpolation onto a finer uniform grid. The fourth quadrant plot 
collects results from both the 1d and 3d experiments. The black 
circles correspond to the $L_\infty$ errors from the 
1d code shown in the other 
plots (and are taken over all fields). The red squares are 
$L_\infty$ errors from the 3d experiment (and, again, are taken
over all fields). Note that these errors have been computed in 
3d. In both cases we see clean fourth-order convergence.
\begin{figure}[t]
\begin{center}
\includegraphics[width=11.0cm]{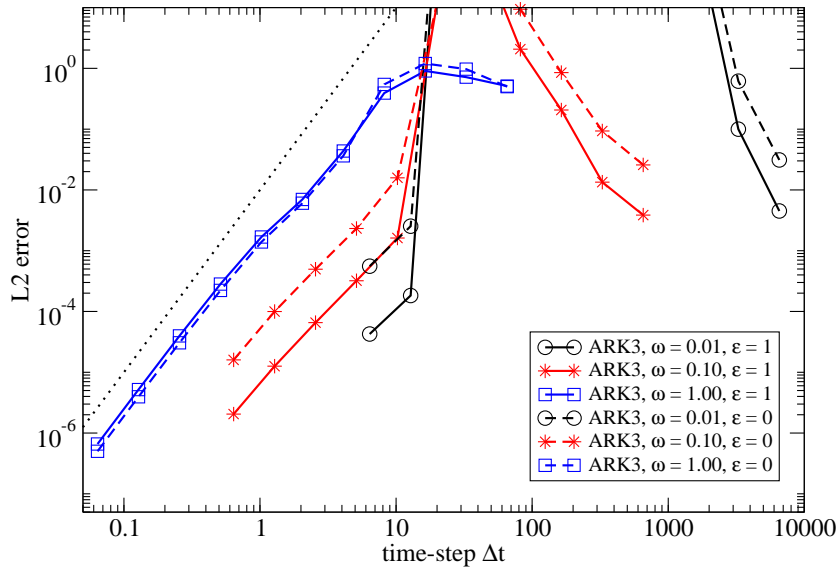}
\end{center}
\caption{Performance of ARK3 for case (ii) IMEX splitting.}
\label{fig:ARK3case_ii}
\end{figure}

\subsection{Model problem on a black hole} \label{subsec:SchwarzModel}

We now consider a problem for which temporal variations occur on
time-scales much longer than the Courant limit for an explicit 
time-stepper.  This model mimics the binary black hole configuration,
providing a testing ground for our IMEX methods.  Our model problem on
a single $M=1$ black hole is set up as follows.  For the solution we
adopt the {\em Ansatz} 
\begin{equation}\label{eq:Psi0}
\psi_0(t,x,y,z) = \cos(\omega t) f(r)
\text{Re}\big[Y_{21}(\theta,\phi)\big],
\end{equation}
where $f(r)$ is a radial
profile.  This profile is chosen as a polynomial of degree $2q$,
\begin{equation}\label{eq:fprofile_model}
f(r) = 
\left(\frac{4}{r_2-r_1}\right)^q\big[(r-r_1)(r_2-r)\big]^q,
\quad r_2 > r_1,
\end{equation}
truncated so as to vanish whenever $r$ lies outside the interval
$[r_1, r_2]$.  We typically choose $q=5$, $r_1 = -1$, and $r_2 =
11.9$.  The computational domain covers radii $r\in[r_\mathrm{min},
  r_\mathrm{max}]=[1.9,11.9] \subset [r_1,r_2]$.  Note that
$r_\mathrm{min}=1.9$ is somewhat inside the black hole horizon,
$r=2$ (recall that $M=1$), and we therefore never apply a boundary 
condition at
$r_\mathrm{min}$.  Nevertheless, we have chosen the support of $f$
such that $f$ is non-zero at the inner edge $r_\mathrm{min}$ of the
computational domain.  However, $f$ does vanish at the outer boundary
$r_\mathrm{max}$, a necessary requirement for avoiding
boundary-driven temporal order reduction \cite{Abarbanel1996}. We
substitute the chosen $\psi_0$ into Eq.~(\ref{eq:ScalarWave2ndOrder})
to compute the source $S$. We furthermore initialize the initial
conditions for $\psi$, $\Pi$ and $\Phi_k$ with the {\em Ansatz}
$\psi_0$.   The $\omega$ value determines
  the time-scale of the temporal variations, and we present results 
for three values,
  $\omega=1, 0.1, 0.01$.  For $\omega=1$, temporal and spatial
  scales are comparable, whereas for $\omega=0.01$, temporal variations are
  vastly slower, so that explicit time-steppers will be limited by the
  Courant condition. 
\begin{figure}[t]
\begin{center}
\includegraphics[width=11.0cm]{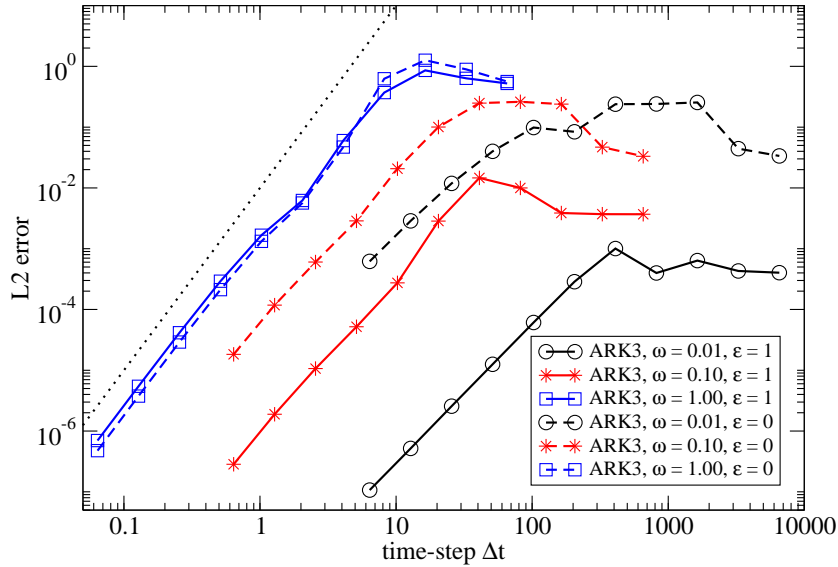}
\end{center}
\caption{Performance of ARK3 for case (iii) IMEX splitting.}
\label{fig:ARK3case_iii}
\end{figure}

Our radial expansions use
Chebyshev polynomials $T_k(X)$ as
basis functions, where we map $X\in[-1,1]$ to $r\in[r_{\rm 
min},r_\mathrm{max}]$ via 
\begin{equation}\label{eq:ExpMap}
r(X) = A e^{B X}+C,
\end{equation}
with $C=-2$ and parameters $A$ and $B$ chosen such that 
$r(-1)=r_\mathrm{min}$, $r(+1)=r_\mathrm{max}$. The 
mapping~(\ref{eq:ExpMap}) serves two purposes: First, it 
increases resolution close to the
black hole, resulting in a somewhat faster convergence rate 
for the spectral representation of the 
Schwarzschild background, Eq.~(\ref{eq:KerrSchild}).  
Second, through this 
mapping the expansion of the radial profile $f(r)$ in 
Chebyshev polynomials acquires non-zero power in {\em all} 
radial modes. 
In contrast, the linear map $r(X) = X$ would result in only 
the first $2q+1$ Chebyshev polynomials being excited.

With either ARK3 or ARK4 and one of the considered IMEX
splittings, our experiment is to evolve the initial data 
specifying this solution, assuming that the field
equations include the exact forcing function $S(t,x,y,z)$.  
We evolve for about 10 oscillation periods, to final time 
 $T_{\mathrm{final}} = 65.536/\omega$. 
We have chosen $N_r = 25$ radial collocation 
points, and the angular grid is determined by top azimuthal index 
$\ell_\mathrm{max} = 5$.  
\begin{figure}
\begin{center}
\includegraphics[width=11.0cm]{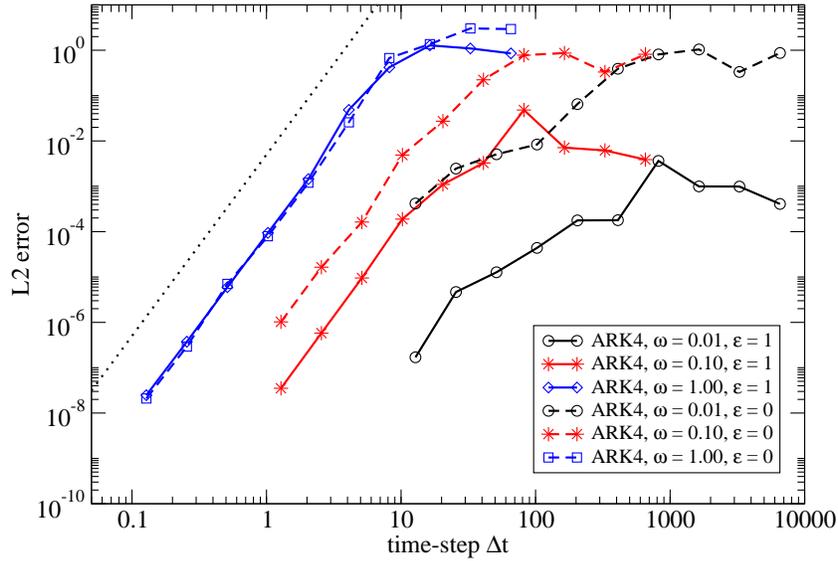}
\end{center}
\caption{Performance of ARK4 for case (iii) IMEX splitting.
The dotted line corresponds to exact fourth-order convergence.}
\label{fig:ARK4case_iii}
\end{figure}

To examine the influence of the IMEX splitting on both long and short 
evolutions, we perform numerical runs with ARK3 for each of the 
aforementioned cases (i), (ii), and (iii). The results are collected 
in Figs.~\ref{fig:ARK3case_i}, \ref{fig:ARK3case_ii}, and 
\ref{fig:ARK3case_iii}. Some of the errors in Figs.~\ref{fig:ARK3case_i}
and \ref{fig:ARK3case_ii} correspond to blowup and fall outside of the 
plot range. Comparing these plots, we notice that for $\omega = 1$ 
short-time runs the accuracy is insensitive to the choice of splitting. 
However, for the small-$\omega$, longer-time runs, 
the fully implicit case (iii) is advantageous in the following sense.
As $\omega$ is reduced by a factor
  of $10$ (from $1$ to $0.1$, and then to $0.01$), this splitting
  allows for a corresponding increase of the time-step $\Delta t$ by 
the same factor of $10$ {\em without} loss of accuracy.
Fig.~\ref{fig:ARK4case_iii} shows results for the same case (iii) 
experiment, but with ARK4 rather than ARK3 used for time integration. 
The fully implicit scenario, that is case (iii) and $\epsilon = 1$, 
corresponds to evolving solely with the $L$-stable ESDIRK component 
of the ARK algorithm \cite{KC-ARK}.  

When this same oscillating multipole problem is evolved by the
explicit fifth-order Cash-Karp scheme \cite{CashKarp1990} the
Courant limit is about $\Delta t_{\rm CFL} \simeq 0.235$ {\em
  independent} of $\omega$.  For all splittings and for all
  $\omega$, the IMEX code allows for time-steps one to two orders of
  magnitude larger than the explicit code.  For slow temporal
  variations, $\omega=0.01$, the splitting (iii) with $\epsilon=1$
  allows for a time-step $\Delta t=1000\Delta t_{\rm CFL}$ 
  (i.e.~about 3 time-steps per oscillation period) while
  maintaining an accuracy of about $10^{-3}$.
\begin{figure}
\begin{center}
\includegraphics[width=11.0cm]{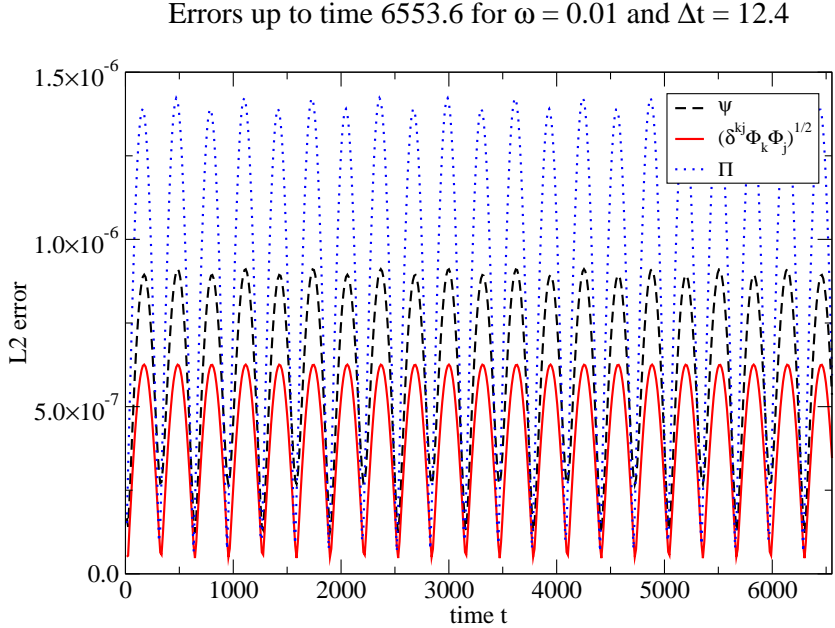}
\end{center}
\caption{Three-dimensional off-center experiment with ARK4.}
\label{fig:offcenterARK4}
\end{figure}

\subsection{Off-center model problem on a 
black hole}\label{subsec:SchwarzModelOffCen}
We now consider the following off-center {\em Ansatz}:
$\psi_0(t,x,y,z) = \cos(\omega t) f(|\mathbf{r}-\mathbf{r}_0|)$, 
where $\omega = 0.01$ and relative to the black hole center 
$\mathbf{r}_0 = (0.5,-0.2,0.3)$. The radial profile is determined as 
above with $r_1 = -1$ and $r_2 = 14$. The numerical domain is 
comprised of three nested spherical shells, each with center 
$\mathbf{c}_0=(-0.08,0.05,-0.06)$. Therefore, each shell is neither a level-$r$ 
surface nor a metric sphere with respect to the background Kerr-Schild 
geometry. Relative to their common center, the shells are determined 
by the radial bounds 1.8, 5.13, 8.47, and 11.8, with coordinate 
radial separations computed using the background Cartesian 
coordinates described before Eq.~(\ref{eq:3metric}). Each shell
has $N_r = 15$ Chebyshev-Lobatto collocation points, with a 
top azimuthal index $\ell_\mathrm{max} = 9$ fixing the angular grid. 
The numerical code expands variables in spherical harmonics centered
on $\mathbf{c}_0$. Because $\mathbf{c}_0\neq\mathbf{r}_0$ all 
spherical-harmonic modes are excited in this experiment.

For this type of exact solution (which involves exact control of a nonzero 
$U^{1-}$ as an inhomogeneous boundary condition), we expect temporal 
order-reduction, a well-known pitfall of exact time-dependent boundary 
conditions \cite{Abarbanel1996}. Therefore, our purpose here is not to 
consider temporal convergence, rather to demonstrate robustness of our 
evolution procedure in a 
setting which mixes several issues at once: an asymmetric solution, 
absence of an inner boundary condition, and multiple domains. While the 
inner boundary lies within the horizon, the coordinate characteristic 
speeds vary spatially across it.
Fig.~\ref{fig:offcenterARK4} depicts long-time error histories 
for all fields using the fully implicit ARK4 time-stepper,
that is case (iii) with $\epsilon = 1$. The plot 
clearly shows the fields' response to the external forcing, with the 
errors continuing to oscillate. At least for linear problems we 
consider here, we believe that our implicit evolutions are 
robustly stable, even in the absence of an inner boundary condition.

\subsection{Model problem with perturbed initial data}
\label{subsec:ModifiedModelProblem}
\begin{figure}[t]
\begin{center}
\includegraphics[width=11.0cm]{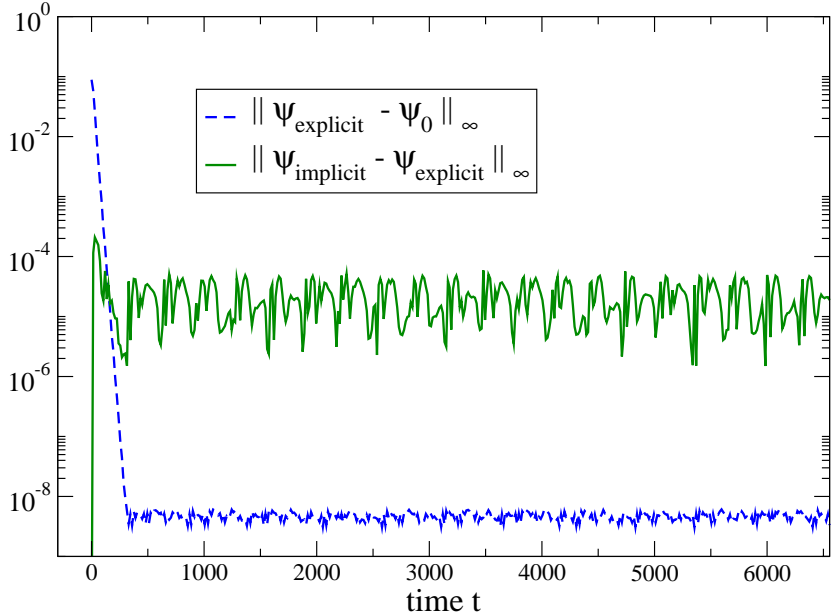}
\end{center}
\caption{{\sc Errors for the modified model experiment.} Note the 
dotted curve giving the deviation of the 
explicit-reference $\psi$ from the {\em Ansatz} $\psi_0$. The 
large deviation at early
times is present since the {\em Ansatz} is not the exact solution.}
\label{fig:AdaptiveErrors}
\end{figure}

  The numerical experiments described in Subsections
  \ref{subsec:SchwarzModel} and \ref{subsec:SchwarzModelOffCen} deal
  with scenarios in which our IMEX integration 
  is simply driven by an external forcing, and 
as a result no secular errors accumulate. That we are
  therefore able to achieve reasonable accuracy for large time-steps
  is perhaps not surprising.  In this section, we provide one further
  test which combines transient, rapidly changing behavior at early
  times with a slowly varying solution at late times.  This test
  mimics start-up effects encountered in binary blackhole simulations, 
  which typically
  exhibit rapid transient behavior at early 
  times when the black holes settle down from imperfect initial data 
  into their quasi-stationary configuration.

We still solve the scalar wave equation with the same source term 
as before,
\begin{equation}
\nabla_\mu \nabla^\mu \psi = S,\qquad S \equiv 
\nabla_\mu \nabla^\mu \psi_0,
\end{equation}
with $\psi_0(t,x^k)$ given in Eq.~(\ref{eq:Psi0}), and here 
with $\omega = 0.01$. However, we now choose initial conditions 
for generating $\psi(t,x^k)$ which are inconsistent 
with those for generating $\psi_0(t,x^k)$. Specifically, we choose
\begin{subequations}
\begin{align}
\psi(0,x^k) &= \psi_0(0,x^k)\\
\Pi(0,x^k) &= \Pi_0(0,x^k)+ G(x^k)\\
\Phi(0,x^k) &= \Phi_0(0,x^k),
\end{align}
\end{subequations}
 where $G(r\nu^k) = 
\exp[-(r-5)^2]\mathrm{Re}[Y_{11}(\theta,\phi)]$ is the 
angularly modulated Gaussian wave packet used 
in Sec.~\ref{subsec:compare}, see Eq.~(\ref{eq:GaussianPacket}). 
Because of the presence of $G(x^k)$, the solution to this evolution 
problem is not simply $\psi(t,x^k)=\psi_0(t,x^k)$, but rather 
there will be an initial deviation. 
For long evolutions, the effect of the Gaussian perturbation 
dies away (due to our dissipative radiation boundary conditions), 
and $\psi(t,x^k) \sim  \psi_0(t,x^k)$ for large $t$. 
For this experiment $t_\mathrm{final} = 6553.6$.

We again work with a single, centered, spherical-shell
  domain and $N_r = 61$, $\ell_\mathrm{max} = 5$.  Rather than the
  mapping (\ref{eq:ExpMap}), now we choose the identity $r(X) = X$.
  This is necessary to fully resolve the Gaussian at early times, or
  else we would require even more radial points.  We will again
  generate a reference numerical solution using an explicit
  time-stepper, in this case Dormand Prince 5 (DP5)
  \cite{numrec_cpp}, against which we will compare an IMEX numerical
  solution obtained with ARK4, choosing case (iii) and $\epsilon = 1$
  so that the evolution is fully implicit. 

A key difference between 
this experiment, and the ones considered in previous subsections, is 
that we now exploit {\em adaptive} time-stepping with {\em dense 
output}. Both DP5 and ARK4 allow for error control and dense output.
The adaptive time-stepping, based on a 
proportional-integral controller described in
\cite{numrec_cpp}, allows the IMEX method
to use small time-steps during the initial transients, and large time-steps
once the transients have died away.
Dense output allows us to conveniently keep track of the error 
history between the explicit-reference and implicit numerical 
solutions. Throughout the course of both the explicit-reference 
and implicit evolutions, we output the solution component $\psi$ 
at all times divisible by 15. For both the explicit-reference and 
implicit evolutions we choose an initial step size $\Delta t = 0.04$, 
with the absolute error tolerance $10^{-5}$.

\begin{figure}[t]
\begin{center}
\includegraphics[width=11.0cm]{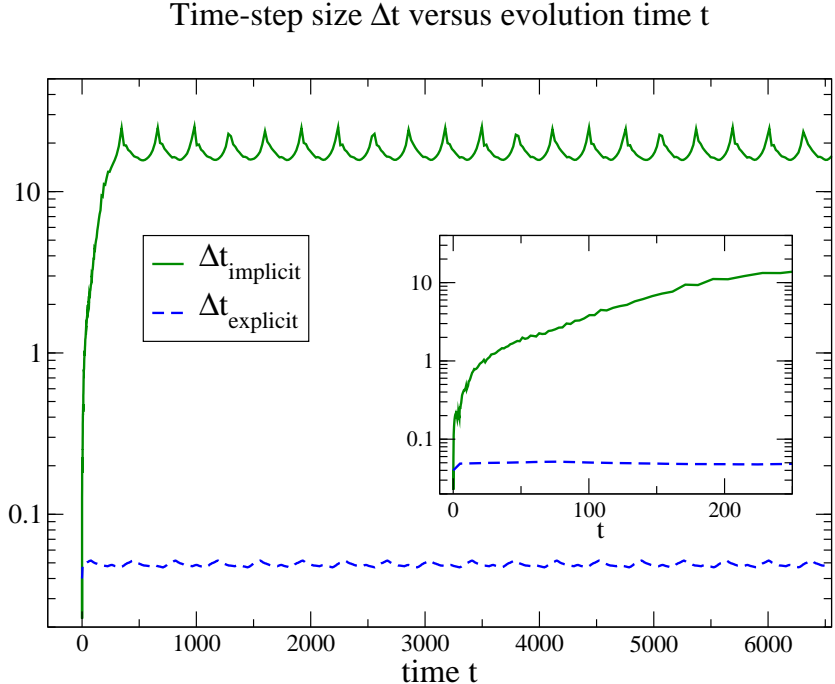}
\end{center}
\caption{{\sc Time-step sizes for explicit-reference and implicit
evolutions.}}
\label{fig:TimeStepSize}
\end{figure}

Figure \ref{fig:AdaptiveErrors} depicts the history of 
the $L_\infty$ difference between the explicit-reference and implicit 
$\psi$, showing that the implicit $\psi$ maintains uniform accuracy 
throughout the evolution. Also depicted in Fig.~\ref{fig:AdaptiveErrors} 
is the deviation of the explicit-reference $\psi$ relative to 
$\psi_0(t,x^k)$. Although $\psi_0(t,x^k)$ is not the exact 
solution, the figure shows that it effectively is for times 
later than $t = 500$. Figure~\ref{fig:TimeStepSize} depicts the 
time-step sizes taken 
throughout the explicit-reference and implicit evolutions. As seen 
in Fig.~\ref{fig:TimeStepSize}, the DP5 evolution essentially runs at 
a fixed time-step $\Delta t \simeq 0.05$ near the Courant limit, and 
in fact this evolution took over $1.3 \times 10^5$ time-steps. By 
contrast, only 475 time-steps were taken during the implicit 
evolution. For the implicit evolution, the time-step size starts off 
small and remains near $0.1$ while the Gaussian pulse propogates off 
the domain. However, at later times the step size dynamically relaxes 
to a time-step of order $\simeq 20$.

\section{Conclusion}\label{sec:conclusion}

As noted in the introduction, implicit or IMEX time-stepping 
is bound to offer a more efficient means of carrying out BBH 
evolutions, especially evolutions involving black holes with 
markedly unequal masses. In the context of scalar waves on
a single black hole, this paper has analyzed several issues 
pertinent to the eventual use of IMEX methods in actual BBH 
evolutions based on the generalized harmonic system. These 
include the role of constraints, the need for second-order 
implicit solves, and the nature of the IMEX splitting. 

Specifically, we have investigated the  role of a pure outflow 
boundary within the black hole horizon. Consistent with the physics, 
the initial boundary value problem associated with the
hyperbolic system of PDEs does not require a boundary condition 
on the inner boundary. Naively, one would 
expect that a second-order implicit equation, as used in our work, 
would require both outer and inner boundary 
conditions, in disagreement with both the underlying 
physics and hyperbolic PDE.  However, the second-order
equation is singular at the horizon, and 
by requiring that the solution is {\em regular} across the horizon, we have
found that the outer boundary condition alone yields uniqueness.

We have examined
the impact of different IMEX splittings on the evolution of a 
wave equation, finding that the performance of our IMEX schemes 
depends sensitively on the precise 
splitting choice [cf.~Figs.~\ref{fig:ARK3case_i} 
to~\ref{fig:ARK3case_iii}].  Only the fully implicit 
choice [case (iii), $\epsilon=1$] allows for time-steps proportional to 
the temporal time-scale, i.e.~$\Delta t\propto 1/\omega$, 
while retaining accuracy independent of $\omega$.
We explain this result as follows. For small 
$\omega$, the right-hand sides of the 
evolution equations~(\ref{eq:FirstOrderScalarWave}) are
$\mathcal{O}(\omega)$ by construction.  However, 
individual terms in the 
expressions are $\mathcal{O}(1)$, because $\psi$, $\Pi$ and $\Phi_k$ 
are all $\mathcal{O}(1)$.  Only the 
{\em sum} of all terms is $\mathcal{O}(\omega)$.  
Consequently, for cases (i) and (ii), and for 
case (iii) with $\epsilon=0$, both the implicit and explicit sectors are large, 
while their sum is small. That is, $\boldsymbol{f}^I$ and $\boldsymbol{f}^E$ 
are $\mathcal{O}(1)$, while 
$\boldsymbol{f}^E+\boldsymbol{f}^I=\mathcal{O}(\omega)$.
Therefore, in the ARK scheme, both the 
implicit  and the explicit sectors involve large drivings, which 
seems to degrade performance. In the general case, we conjecture that  
the IMEX splitting should ideally ensure
 that both $\boldsymbol{f}^E$ and 
$\boldsymbol{f}^I$ remain small. 
  
These observations suggest that a fully 
implicit treatment of the GHS equations will afford accurate evolutions
with very large time-steps. 
However, a corresponding gain in efficiency may well be
offset by the complexity of solving complicated nonlinear implicit
equations.  A more 
workable approach might be to linearize the GHS equations about the 
solution at the current time-step, in order to treat terms with 
constant or linear time-dependence implicitly, and to 
treat terms with quadratic (or higher) time-dependence explicitly.
In this case $\boldsymbol{f}^E$ would be 
$\mathcal{O}(\Delta t^2)$, and perhaps sufficiently small for rather 
large $\Delta t$.  

Another possibility for the
IMEX splitting is particularly promising. Namely, splitting by 
location (or subdomain), as described in 
Ref.~\cite{Kanevskyetal2007} for fluid flow past a nozzle,
a problem for which explicit numerical evolutions are hampered
by boundary induced stiffness.
To understand the idea behind this possibility, consider the 
type of multidomain BBH evolutions now being carried out by the 
Caltech-Cornell collaboration. Such evolutions involve a 
computational domain which is split into about
 60 subdomains (typically spherical shells, cylindrical
shells, and full cylinders with axes). Among these are several 
concentric spherical coordinate shells which enclose each of the 
individual black holes. For either black hole, the innermost of 
these shells contains a topologically spherical apparent horizon. 
As these shells are closest to the black holes where field gradients 
and nonlinearities are the strongest, they require high resolution.
Whence these shells determine the Courant limit for current BBH 
evolutions based on the generalized harmonic system with spectral
methods. 

In those shells nearest the black holes, the splitting we plan to 
investigate would put the local representation of the GHS system 
into the implicit sector, while the equations on all other 
subdomains would be retained in the explicit sector.  The resulting
evolution scheme would still be subject to a (milder) Courant limit 
arising from the grid spacing in those subdomains treated explicitly.  
However, this Courant limit would be independent of the resolution close to
the black holes,  promising efficiency gains 
as the mass ratio increases.   
Implicit equations would need to be solved only in a set of concentric
spherical shells, rather than in a complicated overlapping
domain decomposition,  simplifying preconditioning and improving the 
efficiency of the elliptic solver.  Another reason further 
motivates our interest in an IMEX splitting by location. 
For BBH evolutions based on the GHS system, implementation of 
outer boundary conditions (relevant only for the outermost 
spherical shell enclosing the collection of all inner subdomains) 
involves second derivatives of the physical fields 
\cite{Lindblom2006}. The IMEX splitting we envision would treat 
the outermost spherical shell explicitly, thereby leaving in 
place the current implementation of outer boundary conditions.

  Finally, we point out that for black hole 
  binaries our IMEX time-stepping strategy will only apply in
  {\em co-rotating coordinates}.
  Only in such coordinates does the binary configuration
  appear approximately time-independent, as the black holes 
  remain at the same location in the computational grid. Moreover, the 
  pattern of the emitted gravitational radiation will be almost
  time-independent, varying only on the inspiral time-scale together
  with the orbital frequency and the gravitational wavelength.  This
  restriction is not onerous for our approach, as the {\tt SpEC} code
  already uses co-rotating coordinates within the dual-coordinate
  frame approach developed in Ref.~\cite{Scheel2006}\footnote{In
    this approach {\em inertial-frame} 
    components of tensors are evolved, and these
    components vary on the orbital time-scale.  
    Accuracy considerations will then limit the achievable 
    time-step to the order of the orbital time-scale,
    rather than the longer inspiral time-scale.  The 
    orbital time-scale is still a tremendous improvement over 
    current time-step limitations.   }.

\section*{Acknowledgments}
We would like to thank Thomas Hagstrom, Lawrence Kidder, Lee Lindblom,
Geoffrey Lovelace, Michael Minion, Mark Scheel and Saul
Teukolsky for useful discussions.  Most of the numerical simulations
presented here were performed using the Spectral Einstein Code (SpEC)
developed at Caltech and Cornell primarily by Larry Kidder, H.~P., and
Mark Scheel.  We also thank the referee for comments which
led to the experiment considered in Subsection
\ref{subsec:ModifiedModelProblem}. Revisions were carried out
after S.~L.~had moved to UNM.
This work was supported by grants from the Sherman Fairchild
Foundation and from the Brinson Foundation to Caltech; by grants
DMS 0554377 and DARPA/AFOSR FA9550-05-1-0108 to Brown University;
and by NSF grants PHY-0601459, PHY-0652995 and NASA grant
NNG05GG52G to Caltech.

\appendix


\section{Boundary conditions for first-order 
implicit equations}\label{sec:BCSfirstImpEQNS}
This appendix considers a first-order system similar to 
all three of our first-order implicit systems [Eqs.~(\ref{eq:case1}), 
(\ref{eq:case2}), and (\ref{eq:case3})], showing that the 
system requires 5 boundary conditions. Any of our original 
systems [Eqs.~(\ref{eq:case1}), (\ref{eq:case2}), or 
(\ref{eq:case3})] could be analyzed in a similar fashion, 
although doing so would require a mode decomposition based 
on vector spherical harmonics. Here we use simple Fourier 
series. Consider the system
\begin{subequations}\label{eq:PeriodicIMEX}
\begin{align}
\label{eq:PeriodicIMEXa}
\psi - \alpha (V^x \partial_x\psi - \Pi) 
& = B_\psi
\\
\label{eq:PerodicIMEXb}
\Pi-\alpha (V^x \partial_x \Pi - \partial_k\Phi_k) 
& = B_\Pi
\\
\label{eq:PeriodicIMEXc}
\Phi_k - \alpha (V^x \partial_x \Phi_k - \partial_k\Pi) 
& = B_{\Phi_k}.
\end{align}
\end{subequations}
where the constant shift $V^x$ obeys $0 < V^x < 1$. Take the 
rectangular computational domain to be periodic in the $y$ and
$z$ directions, and lying between $x=0$ and $x=1$. Fourier 
transformation in $y$ and $z$ yields the transformed 
system 
\begin{subequations}
\begin{align}
\label{eq:Periodicftpsi}
\hat{\psi} - \alpha (V^x \partial_x\hat{\psi} 
- \hat{\Pi}) 
& = \hat{B}_\psi
\\
\label{eq:PeriodicftPi}
\hat{\Pi}-\alpha (V^x \partial_x \hat{\Pi} 
- \partial_x\hat{\Phi}_1
- i k_2 \hat{\Phi}_2 - i k_3\hat{\Phi}_3) 
& = \hat{B}_\Pi
\\
\label{eq:PeriodicftPhix}
\hat{\Phi}_1 - \alpha (V^x \partial_x \hat{\Phi}_1 
- \partial_x\hat{\Pi}) 
& = \hat{B}_{\Phi_1}
\\
\label{eq:PeriodicftPhiy}
\hat{\Phi}_2 - \alpha (V^x \partial_x \hat{\Phi}_2 - i k_2\hat{\Pi}) 
& = \hat{B}_{\Phi_2}
\\
\label{eq:PeriodicftPhiz}
\hat{\Phi}_3 - \alpha (V^x \partial_x \hat{\Phi}_3 - i k_3 \hat{\Pi}) &=
\hat{B}_{\Phi_3},
\end{align}
\end{subequations}
where $k_2$ and $k_3$ are the integers dual to $y$ and $z$. All of
the hatted variables should also carry these integer indices,
e.~g.~$\hat{\Pi} = \hat{\Pi}(k_2,k_3)$, but we suppress this
dependence throughout. We replace equations (\ref{eq:PeriodicftPi}) 
and (\ref{eq:PeriodicftPhix}) with the lightlike 
combinations
\begin{align}
\label{eq:PeriodicUplus}
\hat{U}{}^+ - \alpha \big[(V^x-1) 
\partial_x \hat{U}{}^+_x - i k_2 \hat{\Phi}_2 - i 
k_3\hat{\Phi}_3\big] & = \hat{B}_\Pi + \hat{B}_{\Phi_1}
\\
\label{eq:PeriodicUminus}
\hat{U}{}^- - \alpha \big[(V^x+1) \partial_x 
\hat{U}{}^-_x - i k_2 \hat{\Phi}_2 - i
k_3\hat{\Phi}_3\big] & = \hat{B}_\Pi - \hat{B}_{\Phi_1},
\end{align}
thereby arriving at the following inhomogeneous linear 
system:
\begin{equation}\label{eq:Periodicftsystem}
\frac{d}{dx}
\left(\begin{array}{c}
\hat{\psi}\\
\\
\hat{U}{}^+\\
\\
\hat{U}{}^-\\
\\
\hat{\Phi}_2\\
\\
\hat{\Phi}_3
\end{array}\right)
= 
\left(\begin{array}{ccccc}
\frac{1}{\alpha V^x} & \frac{1}{2V^x} & \frac{1}{2V^x} & 0 & 0
\\
& & & &\\
 0 & \frac{1}{\alpha(V^x-1)} & 0 
& \frac{ik_2}{(V^x-1)} 
&
\frac{ik_3}{(V^x-1)}
\\
& & & & \\
0 & 0 & \frac{1}{\alpha(V^x+1)}
& \frac{ik_2}{(V^x+1)}
&
\frac{ik_3}{(V^x+1)}
\\
& & & & \\
0 & \frac{ik_2}{2V^x} & \frac{ik_2}{2V^x} & \frac{1}{\alpha V^x} & 0 
\\
& & & & \\
0 & \frac{ik_3}{2V^x} & \frac{ik_3}{2V^x} & 0 & \frac{1}{\alpha V^x} 
\end{array}\right)
\left(\begin{array}{c}
\hat{\psi}\\
\\
\hat{U}{}^+\\
\\
\hat{U}{}^-\\
\\
\hat{\Phi}_2\\
\\
\hat{\Phi}_3
\end{array}\right)
-
\left(\begin{array}{c}
\frac{\hat{B}_\psi}{\alpha V^x}
\\
\\
\frac{\hat{B}_{U^+}}{\alpha(V^x-1)}
\\
\\
\frac{\hat{B}_{U^-}}{\alpha(V^x+1)}
\\
\\
\frac{\hat{B}_{\Phi_2}}{\alpha V^x}
\\
\\
\frac{\hat{B}_{\Phi_3}}{\alpha V^x}
\end{array}\right).
\end{equation}
With $Q = \sqrt{
 \alpha^2 + \alpha^4 |\mathbf{k}|^2[1 - (V^x)^2]}$ and
$|\mathbf{k}|^2 = k_2^2 + k_3^2$, the eigenvalues of the 
coefficient matrix $\mathbf{A}$ are
\begin{equation}
\label{eq:Periodicevalues}
\lambda_1 = \lambda_2 = \lambda_3 = (\alpha V^x)^{-1},
\qquad \lambda_{4,5} = 
\frac{-\alpha V^x \pm Q}{\alpha^2[1- (V^x)^2]}.
\end{equation}
The corresponding eigenvectors are
\begin{equation}
\label{eq:Periodicevectors}
\boldsymbol{v}_1 =
\left(\begin{array}{c}
1\\
0\\
0\\
0\\
0
\end{array}\right),\;
\boldsymbol{v}_2 =
\left(\begin{array}{c}
0\\
-i\alpha k_2 V^x\\
i\alpha k_2 V^x\\
1\\
0
\end{array}\right),
\;	
\boldsymbol{v}_3 =
\left(\begin{array}{c}
0\\
-i\alpha k_3 V^x\\
i\alpha k_3 V^x\\
0\\
1
\end{array}\right),
\;
\boldsymbol{v}_{4,5} =
\left(\begin{array}{c}
-\alpha^2
\\
(\alpha \mp Q)/(1-V^x)
\\
(\alpha \pm Q)/(1+V^x)
\\
-i\alpha^2 k_2
\\
-i\alpha^2 k_3
\end{array}\right).
\end{equation}
The $(k_2,k_3) = (0,0)$ limits of Eqs.~(\ref{eq:Periodicevalues})
and (\ref{eq:Periodicevectors}) are easily computed with the 
result $Q \sim \alpha$, $|\mathbf{k}| \rightarrow 0^+$. The 
results agree with those obtained by first setting $(k_2,k_3) 
= (0,0)$ in (\ref{eq:Periodicftsystem}), and then performing the 
eigen-decomposition. Notice that the $|\mathbf{k}| = 0$
eigenvalues, which happen to be the diagonal entries of the 
coefficient matrix $\mathbf{A}$ in (\ref{eq:Periodicftsystem}), 
are such that $(\alpha\lambda_q)^{-1}$ for $q = 1,\ldots,5$ are 
the characteristic speeds of the corresponding hyperbolic system. 

The eigenvectors (\ref{eq:Periodicevectors}) are not mutually 
orthogonal; however,
\begin{equation}
\det\big[
\boldsymbol{v}_1,
\boldsymbol{v}_2,
\boldsymbol{v}_3,
\boldsymbol{v}_4,
\boldsymbol{v}_5\big] 
= -\frac{4\alpha \big[1 - \alpha^2 |\mathbf{k}|^2 
(V^x)^2\big]Q}{\big[1-(V^x)^2\big]},
\end{equation}
and the eigensolutions
\begin{equation}
\boldsymbol{y}_q(x) = 
e^{\lambda_q x} \boldsymbol{v}_q,\quad q = 1,2,\ldots,5
\end{equation}
form a fundamental set of solutions. Defining
\begin{equation}
\boldsymbol{\Psi}(x) = \big[
\boldsymbol{y}_1(x),
\boldsymbol{y}_2(x),
\boldsymbol{y}_3(x),
\boldsymbol{y}_4(x),
\boldsymbol{y}_5(x)\big]
\end{equation}
and viewing the system (\ref{eq:Periodicftsystem}) as
\begin{equation}
\frac{d}{dx}\boldsymbol{y}(x) = \mathbf{A}\boldsymbol{y}(x) + \mathbf{g}(x),
\end{equation}
we can now write down the general solution:
\begin{equation}
\boldsymbol{y}(x) = \boldsymbol{\Psi}(x)\mathbf{c} +
\boldsymbol{\Psi}(x) \int^x_{x_0} \boldsymbol{\Psi}^{-1}(\xi)
\mathbf{g}(\xi)d\xi,
\end{equation}
where $x_0$ is any point on the interval $(0,1)$. The five 
components $c_q$ of $\mathbf{c}$ correspond to five boundary 
conditions. The following recipe for fixing these
components agrees with the convention for control of
incoming fields in the corresponding evolution 
initial-boundary-value
problem. The exponentials $e^{\lambda_q x}$ for $q=1,2,3,4$ all
blow up as $x\rightarrow\infty$, whereas $e^{\lambda_5 x}$
decays in the same limit. We want to fix the 
eigensolutions $\boldsymbol{y}_q(x)$ for $q=1,2,3,4$ 
(associated with blowing-up exponentials) at $x=1$, and the 
eigensolution $\boldsymbol{y}_5(x)$ (associated with the sole 
decaying exponential) at $x=0$. We assume that $(k_2,k_3)$ is 
small, so that the eigenvectors 
$\boldsymbol{v}_1$,$\boldsymbol{v}_2$, 
$\boldsymbol{v}_3$, and $\boldsymbol{v}_4$ are combinations of 
the fields $\hat{\psi}$, $\hat{U}^-$, $\hat{\Phi}_2$, and 
$\hat{\Phi}_3$. Therefore, we fix these fields at $x=1$. Also 
for small $(k_2,k_3)$, $\boldsymbol{v}_5$ is approximately 
proportional to the fields $\hat{\psi}$ (already fixed at $x=1$) 
and $\hat{U}^+$ which would be the $\hat{U}^-$ field relative
to the outward-pointing unit normal $-d/dx$ at $x=0$. Finally
then, we fix $\hat{U}^+$ at $x=0$.  

\section{Singular boundary value problem}\label{sec:SBVP}

In this appendix we consider the general solution to the 
second-order equation (\ref{eq:appendixeqn}) for the case of 
the Schwarzschild geometry with respect to Kerr-Schild 
coordinates. Provided that the radial location 
$r = r_\mathrm{min}$ of inner boundary satisfies $r_\mathrm{min} 
\leq 2M$, we show that a regular 
(that is, nonsingular) solution to the equation is uniquely
determined by one free constant. We conclude that a single 
outer boundary conditions suffices to determine a regular 
solution to the equation. 

Our analysis assumes that both $\mathcal{Q}(r)$ and $\mathcal{P}(r)$ 
in Eq.~(\ref{eq:appendixeqn}) are smooth on the radial domain, which 
is easily checked for all cases. We further note that
$\alpha^{-1}\mathcal{P}(2M) > 1$, where $\mathcal{P}(2M) = 
4M\mathcal{S}(2M) + 2\alpha$. Let us verify that this last 
inequality holds for the considered cases. For 
(\ref{eq:smallelliptic}) and case (i) we have
\begin{equation}
\mathcal{S}(r) = 2 \shift^r,
\end{equation}
and so $\alpha^{-1}\mathcal{P}(2M) = 2 + 4M\alpha^{-1} > 1$
by (\ref{eq:geometry}).
For (\ref{eq:mediumelliptic}) and case (ii), we find
\begin{equation}
\mathcal{S}(r) = 2 \shift^r + \alpha \big[(\shift^r)^2 N^{-1}N'
- \shift^r (\shift^r)'\big].
\end{equation}
Calculations with (\ref{eq:geometry}) then show that
$\mathcal{S}(2M) = 1 +\frac{3}{8}\alpha/(4M)$, whence
$\alpha^{-1}\mathcal{P}(2M) = \frac{19}{8} + 4M \alpha^{-1} > 1$.
Finally, we consider
(\ref{eq:largeelliptic})
and
case (iii), with our earlier calculations giving
\begin{equation}
\mathcal{S}(r) =
2\shift^r
+\alpha \left[
(\shift^r)^2 N^{-1}N'
- \shift^r (\shift^r)'
-N^2 (J^k\partial_k r)
- \shift^r NK
\right].
\end{equation}
With the formulas listed in (\ref{eq:geometry}) and
(\ref{eq:JkandEnnKay}), we find $\mathcal{S}(2M) = 1 - \alpha/(4M)$,
implying as claimed that $\alpha^{-1}\mathcal{P}(2M) =
1 + 4M\alpha^{-1} > 1$. 
Our argument is completed with the following:

\noindent
{\sc Lemma:} Consider the {\sc ode}
\begin{equation}
\mathcal{Q}(r)
w + \alpha\mathcal{P}(r)w'
+\alpha^2(r-2M)w''
= h(r),
\label{eq:maineqn}
\end{equation}
here taken on the $r$-interval $(2M,r_\mathrm{max})$. Assume
that $\mathcal{Q}(r)$, $\mathcal{P}(r)$, and $h(r)$ are 
smooth on an open interval larger than 
$(2M,r_\mathrm{max})$. Moreover, assume that 
$\alpha^{-1}\mathcal{P}(2M) > 1$, also with $\alpha > 0$.
Express the general solution as
\begin{equation}
w(r) = c_1 w_1(r) + c_2 w_2(r) + w_P(r),
\end{equation}
where $w_1(r)$ and $w_2(r)$ are solutions to the 
homogeneous equation (that is, for $h(r) = 0$), and 
$w_P(r)$ is a particular solution. Then we may arrange 
for $w_1(r)$ and $w_P(r)$ to be regular as
$r\rightarrow 2M^+$, with $w_2(r)$ 
singular and obeying
\begin{equation}
w_2(r) \sim (r-2M)^{1-1/(\alpha\kappa)},
\end{equation}
again as $r\rightarrow 2M^+$. Here $\kappa = 
1/\mathcal{P}(2M)$, and $1 - 1/(\alpha\kappa) < 0$
by assumption. The second solution must therefore exhibit a 
blowing-up (likely also branch) singularity at $r = 2M$. 

We begin the proof of the lemma by examining the homogeneous
equation. Taken in standard form, that equation is
\begin{equation}
w'' + P(r) w' + Q(r) w = 0,
\end{equation}
where
\begin{equation}
P(r) = \frac{1}{\alpha} 
\frac{\mathcal{P}(r)}{r-2M},\qquad
Q(r) = \frac{1}{\alpha^2} 
\frac{\mathcal{Q}(r)}{r-2M}.
\end{equation}
Seeking solutions of Frobenius type, we then consider the indicial
equation
\begin{equation}
\lambda(\lambda - 1) + \lambda/(\alpha\kappa) = 0.
\end{equation}
Whence the indicial exponents are $\lambda_1 = 0,
\lambda_2 = 
1-1/(\alpha\kappa)$, and we may therefore choose solutions 
to the homogeneous problem obeying
\begin{equation}
w_1(r) \sim 1,\qquad w_2(r) \sim (r-2M)^{1-1/(\alpha\kappa)},
\end{equation}
as $r\rightarrow 2M^+$. At $r = 2M$, the 
first solution is analytic, while the second exhibits blow-up,
and likely branch behavior depending on the value of $\alpha\kappa$.

To complete the proof, we follow the method of undetermined coefficients
in order to construct a particular solution with the desired regularity.
For $r > 2M$, an integrating factor for (\ref{eq:maineqn}) is 
\begin{equation}
(r-2M)^{-1+1/(\alpha\kappa)}
\exp\left[\frac{1}{\alpha}\int_{2M}^r
\frac{\mathcal{P}(\xi) - 
\mathcal{P}(2M)}{\xi-2M}\mathrm{d}\xi\right]
= (r-2M)^{-1+1/(\alpha\kappa)}\mu(r),
\end{equation}
where $\mu(2M) = 1$. Using the integrating factor,
we cast (\ref{eq:maineqn}) into the following form:
\begin{equation}
\big[\alpha^2(r-2M)^{1/(\alpha\kappa)}
\mu(r)w'\big]{}' +
(r-2M)^{-1+1/(\alpha\kappa)}
\mu(r)\mathcal{Q}(r)w
= (r-2M)^{-1+1/(\alpha\kappa)}
\mu(r)h(r).
\end{equation}
It then follows on general grounds that
\begin{equation}
W[w_1,w_2](r) = w_1(r) w_2'(r) - w_2(r) w_1'(r)
= \frac{A}{\alpha^2\mu(r)}(r-2M)^{-1/(\alpha\kappa)},
\end{equation}
where the constant $A = \alpha^2 [1 - 1/(\alpha\kappa)]$.
With this result for the Wronskian $W[w_1,w_2](r)$
in hand, we look for a solution
\begin{equation}
w_P(r) = u(r) w_1(r) + v(r) w_2(r),
\end{equation}
subject to the variation-of-parameters {\em Ansatz}
\begin{equation}
u'(r) w_1(r) + v'(r) w_2(r) = 0.
\end{equation}
The needed expressions for $u(r)$ and $v(r)$ are as follows:
\begin{align}
u(r) = &
-\int_{b}^r A^{-1} w_2(\xi)
(\xi-2M)^{-1+1/(\alpha\kappa)}
\mu(\xi)h(\xi)\mathrm{d}\xi,
\label{eq:uintegral}\\
v(r) = &
\int_{2M}^r A^{-1} w_1(\xi)
(\xi-2M)^{-1+1/(\alpha\kappa)}
\mu(\xi)h(\xi)\mathrm{d}\xi.
\label{eq:vintegral}
\end{align}
More compactly, we may write
\begin{equation}
w_P(r) = \int_{2M}^b 
G(r,\xi)\mu(\xi)(\xi-2M)^{-1+1/(\alpha\kappa)} 
h(\xi)\mathrm{d}\xi,
\end{equation}
in terms of the Green's function
\begin{equation}
G(r,\xi) = 
\left\{\begin{array}{lcl}
                  A^{-1}w_1(r) w_2(\xi) & \mathrm{for} & 2M 
\leq r \leq \xi \leq b\\
A^{-1}w_1(\xi) w_2(r) & \mathrm{for} & 2M 
\leq \xi \leq r \leq b.
                  \end{array}\right.
\end{equation}

Finally, to verify that, as constructed, $w_P(r)$ remains 
regular 
as $r\rightarrow 2M^+$, we establish in the same limit that
\begin{equation}
u(r) \sim K_1,\qquad v(r) \sim K_2 (r-2M)^{1/(\alpha\kappa)},
\end{equation}
for constants $K_1 = u(2M)$ and  $K_2 = \alpha\kappa 
h(2M)$. The first asymptotic statement follows
from the observation that the integrand in (\ref{eq:uintegral}) is
integrable at $r = 2M$. To get the result for $v(r)$, we
use 
\begin{align}
v'(r) & = 
A^{-1} w_1(r)
(r-2M)^{-1+1/(\alpha\kappa)}
\mu(r)h(r)
\nonumber \\
& = A^{-1} w_1(2M) 
\mu(2M)h(2M)(r-2M)^{-1+1/(\alpha\kappa)}
+ O\big((r-2M)^{1/(\alpha\kappa)}\big),
\end{align}
along with $w_1(2M) = 1 = \mu(2M)$. Taken all 
together, we have shown that
\begin{equation}
w_P(r) \sim K_1 + K_2 (r-2M),
\end{equation}
as $r\rightarrow 2M^+$. Whence the lemma has been proved. 
$\Box$

\section{Implicit constraint equations}\label{sec:impC}

For the Schwarzschild example with line-element 
(\ref{eq:KerrSchild}), this appendix further examines 
Eq.~(\ref{eq:constraintimp}). To obtain an orthonormal 
spatial triad, we complete the radial vector $n^k = L^{-1} \nu^k$ 
defined just before Eq.~(\ref{eq:3metric}) with the standard 
angular directions
\begin{equation}
e_{\boldsymbol{\theta}}{}^k
= (\cos\theta\cos\phi,\cos\theta\sin\phi,-\sin\theta),
\quad e_{\boldsymbol{\phi}}{}^k = 
(-\sin\phi,\cos\phi,0).
\end{equation} 
In terms of the triad, we have
\begin{equation}
\mathcal{C}_k = n_k L^{-1} \mathcal{C}_\nu 
+ e_{\boldsymbol{\theta}k} \mathcal{C}_{\boldsymbol{\theta}} 
+  e_{\boldsymbol{\phi}k} \mathcal{C}_{\boldsymbol{\phi}},
\label{eq:CartesianCk}
\end{equation}
where $\mathcal{C}_\nu 
= \nu^k \mathcal{C}_k$,
$\mathcal{C}_{\boldsymbol{\theta}} 
= e_{\boldsymbol{\theta}}{}^k \mathcal{C}_k$,
and $\mathcal{C}_{\boldsymbol{\phi}} 
= e_{\boldsymbol{\phi}}{}^k \mathcal{C}_k$.
Contraction of (\ref{eq:constraintimp}) on $\nu^k$ yields the 
equation
\begin{equation}
\mathcal{C}_\nu - \alpha \big(\shift^r \mathcal{C}_\nu\big)' 
= B_\psi'-\nu^k B_{\Phi_k},
\label{eq:nuconstraintimp}
\end{equation}
with the prime denoting radial differentiation. Likewise, 
contraction of (\ref{eq:constraintimp}) on 
$e_{\boldsymbol{\theta}}{}^k$ yields
\begin{equation}
\mathcal{C}_{\boldsymbol{\theta}} 
- \alpha \big(\shift^r 
\mathcal{C}_{\boldsymbol{\theta}}'
+ r^{-1} \mathcal{C}_{\boldsymbol{\theta}} 
\shift^r\big) =
e_{\boldsymbol{\theta}}{}^k\big(\partial_k B_\psi - 
B_{\Phi_k}\big),
\label{eq:thetaconstraintimp}
\end{equation}
where in reaching this equation we have used
$\shift^k = \shift^r \nu^k$ and 
$e_{\boldsymbol{\theta}}{}^j\partial_j\nu^k 
= r^{-1}e_{\boldsymbol{\theta}}{}^k$.
Similar manipulations establish that
\begin{equation}
\mathcal{C}_{\boldsymbol{\phi}} 
- \alpha \big(\shift^r \mathcal{C}_{\boldsymbol{\phi}}'
+ r^{-1} \mathcal{C}_{\boldsymbol{\phi}} \shift^r\big) =
e_{\boldsymbol{\phi}}{}^k
\big(\partial_k B_\psi - B_{\Phi_k}\big).
\label{eq:phiconstraintimp}
\end{equation}
Since $\shift^r > 0$ over the whole spherical shell, we may radially integrate 
(\ref{eq:nuconstraintimp},\ref{eq:thetaconstraintimp},\ref{eq:phiconstraintimp}) 
inward from the outer boundary $\Bouter$, provided 
$\mathcal{C}_\nu|_\Bouter$, $\mathcal{C}_{\boldsymbol{\theta}} |_\Bouter$,
and $\mathcal{C}_{\boldsymbol{\phi}}|_\Bouter$ are specified.
We can then recover the Cartesian components $\mathcal{C}_k$ with 
(\ref{eq:CartesianCk}). In principle, these components could be 
incorporated into the $\mathcal{G}$ source in 
Eq.~(\ref{eq:generalelliptic}).


\begin{thebibliography}{99}

\bibitem{Pretorius2005}
F.~Pretorius, Evolution of Binary Black-Hole Spacetimes,
Phys.~Rev.~Lett., 95 (2005), 121101 (4 pages).

\bibitem{UTB2006}
M.~Campanelli, C.~O.~Lousto, P.~Marronetti, and Y.~Zlochower,
Accurate Evolutions of Orbiting Black-Hole Binaries without
Excision, Phys.~Rev.~Lett., 96 (2006), 111101 (4 pages).

\bibitem{NASA2006}
J.~G.~Baker, J.~Centrella, D.-I.~Choi, M.~Koppitz, and J.~van Meter,
Gravitational-Wave Extraction from an Inspiraling Configuration of
Merging Black Holes, Phys.~Rev.~Lett., 96 (2006), 111102 (4 pages).

\bibitem{Boyle2007}
M.~Boyle, D.~A.~Brown, L.~E.~Kidder, A.~H.~Mrou{\'e},
H.~P.~Pfeiffer, M.~A.~Scheel, G.~B.~Cook, and S.~A.~Teukolsky,
High-accuracy comparison of numerical relativity simulations with
post-{N}ewtonian expansions, Phys.~Rev.~D, 76 (2007), 124038 (31 pages).

\bibitem{Baker2008}
J.~G.~Baker, W.~D.~Boggs, J.~Centrella, B.~J.~Kelly, S.~T.~McWilliams,
and J.~R.~van Meter, Mergers of nonspinning black-hole binaries:
Gravitational radiation characteristics, Phys.~Rev.~D, 78 (2008),
044046 (25 pages).

\bibitem{Sperhake2008}
J.~A.~Gonzalez, U.~Sperhake, and B.~Br\"ugmann, Black-hole binary
simulations: the mass ratio 10:1,
arXiv:0811.3952v1 [gr-qc] (2008, 10 pages).

\bibitem{KC-ARK}
C.~A.~Kennedy and M.~H.~Carpenter,
Additive Runge-­Kutta schemes for 
convection-­diffusion-­reaction equations,
Appl.~Numer.~Math., 44 (2003), 139-181.

\bibitem{Dutt2000}
A.~Dutt, L.~Greengard, and V.~Rokhlin,
Spectral Deferred Correction Methods for 
Ordinary Differential Equations,
BIT, 40 (2000), 241-266.

\bibitem{Minion2003}
M.~L.~Minion,
Semi-Implicit Spectral Deferred Correction 
Methods for Ordinary Differential Equations,
Commun.~Math.~Sci., 1 (2000), 471-500.

\bibitem{Kanevskyetal2007}
A.~Kanevsky, M.~H.~Carpenter, D.~Gottlieb, and J.~S.~Hesthaven,
Application of implicit-explicit high order Runge-Kutta 
methods to discontinuous-Galerkin schemes,
J.~Comput.~Phys., 225 (2007), 1753-1781.

\bibitem{Pfeiffer2003}
Harald~P.~Pfeiffer, Lawrence~E.~Kidder, Mark~A.~Scheel, 
and Saul~A.~Teukolsky,
A multidomain spectral method for solving elliptic equations,
Comput.~Phys.~Commun., 152 (2003), 253-273.

\bibitem{Pretorius2005c}
F.~Pretorius,
Numerical relativity using a generalized harmonic decomposition,
Class.~Quantum Grav., 22 (2005), 425-452.

\bibitem{Pretorius2006}
F.~Pretorius,
Simulation of binary black hole spacetimes with a harmonic 
evolution scheme, Class.~Quantum Grav., 23 (2006), S529-S552.

\bibitem{Lindblom2006}
L.~Lindblom, M.~A.~Scheel, L.~E.~Kidder, R.~Owen, 
and O.~Rinne, A new generalized harmonic evolution system,
Class.~Quantum Grav., 23 (2006), S447-S462.

\bibitem{HennigAnsorg2008}
J.~Hennig and M.~Ansorg,
A fully pseudospectral scheme for solving singular
hyperbolic equations on conformally compactified
space-times, Journal of Hyperbolic Differential Equations
6, No.~1 (2009), 161-184.

\bibitem{Holst2004}
M.~Holst, L.~Lindblom, R.~Owen, H.~P.~Pfeiffer, 
M.~A.~Scheel, and L.~E. Kidder.
Optimal constraint projection for hyperbolic evolution systems,
Phys.~Rev.~D, 70 (2004), 084017 (17 pages).

\bibitem{Kidder2000a}
L.~E.~Kidder, M.~A.~Scheel, S.~A.~Teukolsky, E.~D.~Carlson, 
and G.~B.~Cook, Black hole evolution by spectral methods,
Phys.~Rev.~D, 62 (2000), 084032 (20 pages).

\bibitem{Scheel2006}
M.~A.~Scheel, H.~P.~Pfeiffer, L.~Lindblom, 
L.~E.~Kidder, O.~Rinne, and S.~A.~Teukolsky,
Solving {E}instein's equations with dual coordinate frames,
Phys.~Rev.~D, 74 (2006), 104006 (13 pages).

\bibitem{CashKarp1990}
J.~R.~Cash and A.~H.~Karp,
A Variable Order Runge-Kutta Method for Initial 
Value Problems with
Rapidly Varying Right-Hand Sides,
ACM Transactions on Mathematical Software, 
16 (1990), 201-222.

\bibitem{Abarbanel1996}
S.~Abarbanel, D.~Gottlieb, and M.~H.~Carpenter,
On the Removal of Boundary Errors Caused by 
Runge-Kutta Integration
of Nonlinear Partial Differential Equations,
SIAM J.~Sci.~Comput., 17 (1996), 777-782.

\bibitem{numrec_cpp} W.~H.~Press, S.~A.~Teukolsky, W.~T.~Vetterling, and 
B.~P.~Flannery, {\em Numerical Recipes: The Art of Scientific 
Computing, Third Edition} (Cambridge University Press, Cambridge, 2007).

\end{thebibliography}
\end{document}